\documentclass[aps,prx,twocolumn,superscriptaddress,10pt]{revtex4-2}
\usepackage[utf8]{inputenc}
\usepackage[english]{babel}
\usepackage[T1]{fontenc}
\usepackage{microtype}
\usepackage[pdftex]{graphicx}
\usepackage{hyperref}
\usepackage{amsmath}
\usepackage{amssymb}
\usepackage{braket}
\usepackage{xcolor}
\usepackage{dcolumn}
\usepackage{bbm}
\usepackage{bbold}
\usepackage[percent]{overpic}
\usepackage[caption=false]{subfig}
\usepackage[capitalize]{cleveref}
\usepackage{float}

\hypersetup{%
    hypertexnames=false,
    colorlinks=true,
    allcolors=blue,
    unicode=true
}

\pdfoutput=1

\newcommand{\includescaledgraphics}[1]{\vcenter{\hbox{\scalebox{0.5}{\includegraphics{#1}}}}}

\newcommand{\phantomsubfloat}[1]{%
    {%
        \captionsetup[subfloat]{farskip=0pt,captionskip=0pt}
        \captionsetup[subfigure]{labelformat=empty}
        \subfloat{#1}
    }
}

\crefname{section}{Sec.}{Sec.}
\crefname{appendix}{App.}{App.}

\DeclareMathOperator{\tr}{Tr}

\newcommand{\id}{\mathbb{1}}

\begin{document}

\title{Trimer states with \texorpdfstring{$\mathbb{Z}_3$}{ℤ₃} topological order in Rydberg atom arrays}
\author{Giacomo Giudice}
\affiliation{Max-Planck-Institute of Quantum Optics, Hans-Kopfermann-Straße 1, 85748 Garching, Germany}
\affiliation{Munich Center for Quantum Science and Technology (MCQST), Schellingstra\ss{}e~4, 80799 M\"{u}nchen, Germany}
\author{Federica Maria Surace}
\affiliation{Department of Physics and Institute for Quantum Information and Matter,
    California Institute of Technology, Pasadena, California 91125, USA}
\author{Hannes Pichler}
\affiliation{Institute for Theoretical Physics, University of Innsbruck, Innsbruck A-6020, Austria}
\affiliation{\mbox{Institute for Quantum Optics and Quantum Information, Austrian Academy of Sciences, Innsbruck A-6020, Austria}}
\author{Giuliano Giudici}
\affiliation{\mbox{Arnold Sommerfeld Center for Theoretical Physics, University of Munich, Theresienstra\ss{}e 37, 80333 M\"{u}nchen, Germany}}
\affiliation{Munich Center for Quantum Science and Technology (MCQST), Schellingstra\ss{}e~4, 80799 M\"{u}nchen, Germany}

\begin{abstract}
    Trimers are defined as two adjacent edges on a graph.
    We study the quantum states obtained as equal-weight superpositions of all trimer coverings of a lattice, with the constraint of having a trimer on each vertex: the so-called trimer resonating-valence-bond (tRVB) states.
    Exploiting their tensor network representation, we show that these states can host $\mathbb{Z}_3$ topological order or can be gapless liquids with $\mathrm{U}(1) \times \mathrm{U}(1)$ local symmetry.
    We prove that this continuous symmetry emerges whenever the lattice can be tripartite such that each trimer covers all the three sublattices.
    In the gapped case, we demonstrate the stability of topological order against dilution of maximal trimer coverings, which is relevant for realistic models where the density of trimers can fluctuate.
    Furthermore, we clarify the connection between gapped tRVB states and $\mathbb{Z}_3$ lattice gauge theories by smoothly connecting the former to the $\mathbb{Z}_3$ toric code, and discuss the non-local excitations on top of tRVB states.
    Finally, we analyze via exact diagonalization the zero-temperature phase diagram of a diluted trimer model on the square lattice and demonstrate that the ground state exhibits topological properties in a narrow region in parameter space.
    We show that a similar model can be implemented in Rydberg atom arrays exploiting the blockade effect.
    We investigate dynamical preparation schemes in this setup and provide a viable route for probing experimentally $\mathbb{Z}_3$ quantum spin liquids.
\end{abstract}

\maketitle

\section{Introduction}

When quantum fluctuations meet classical frustration, exotic strongly correlated states can arise.
Paradigmatic examples are resonating valence bond (RVB) states of hard dimers.
They are defined as equal-weight quantum superpositions of all dimer coverings with one dimer touching each vertex of a lattice.
These many-body states are the ground states of local Hamiltonians with peculiar properties, such as a ground-state degeneracy that depends on the topology of the system and deconfined excitations that come in pairs~\cite{moessner2011}.
When their correlation length is finite, a stable, topologically ordered, quantum phase of matter exists with RVB states as its representatives~\cite{moessner2001}.
This kind of topological order is characterized by a local $\mathbb{Z}_2$ symmetry closely related to the Gauss' law in the gauge theory description of these phases, and it is expected to emerge in dimer models defined on non-bipartite lattices~\cite{moessner2011}.

\begin{figure}[ht]
    \includegraphics[width=\columnwidth]{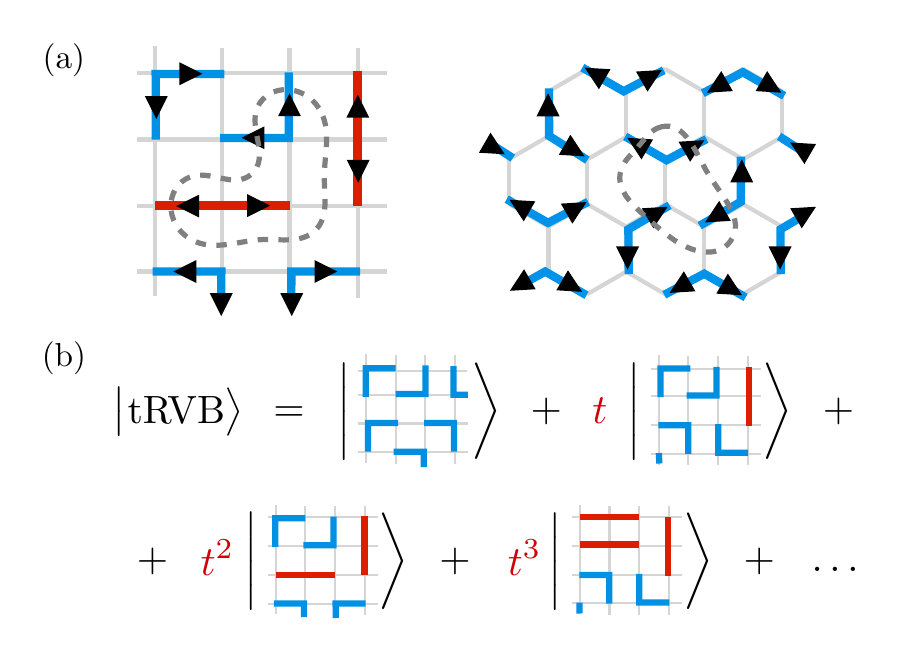}
    \vspace*{-6mm}
    \phantomsubfloat{\label{fig:intro:config}}
    \phantomsubfloat{\label{fig:intro:state}}
    \caption{%
        (a)~Maximal trimer configuration with arrows pointing from the center to the two external vertices of each trimer, on the square and honeycomb lattice.
        The net flux for each vertex is $2 \, \mathrm{mod}\,3$.
        The region enclosed by the dashed line has $N_s=3$ vertices and a net flux $3=2N_s\, \mathrm{mod}\,3$.
        (b)~One-parameter family of tRVB states on the square lattice.
        The parameter $t=\tan \theta$ weights the presence of straight trimers, such that for $\theta = 0$ ($\pi/2$) the tRVB state is solely made of bent (straight) trimers.
    }
    \label{fig:intro}
\end{figure}
In this paper, we consider RVB states of hard trimers (tRVB).
A trimer is an object made up of two nearest-neighbor edges of a lattice that share a common vertex [cf. \cref{fig:intro:config}].
Hard trimers cannot touch each other, such that each vertex of the lattice can be covered by at most one trimer, yielding what we refer to as the trimer constraint.
Maximally-packed trimer configurations are obtained by demanding that exactly one trimer covers each vertex.
These configurations are then promoted to orthogonal quantum states, and the tRVB state is their equal-weight quantum superposition.
On certain lattices, tRVB states are known to be gapped and to possess a form of topological order with emergent $\mathbb{Z}_3$ gauge symmetry~\cite{lee2017,dong2018,jandura2020}, and thus to be good representatives of a quantum phase with $\mathbb{Z}_3$ topological order.
However, a general condition on the lattice geometry for which these states are gapped and have topological character is still lacking, and physical realizations of this phase are little known~\cite{yang2021}.
Here, we identify a necessary condition for having a gapped tRVB state with $\mathbb{Z}_3$ topological order, similar to the condition of non-bipartite lattices for dimer models.
Moreover, we provide a trimer model where signatures of the above-mentioned topological phase are identified.
This model is physically relevant for experimental platforms based on Rydberg atom arrays, as we demonstrate by proposing and analyzing a viable implementation of the trimer constraint in this setup.

Firstly, we focus on the square lattice, where we demonstrate the emergence of the trimer coverings Hilbert space as a particular limit of a $\mathbb{Z}_3$ lattice gauge theory (LGT).
We introduce a tensor network (TN)~\cite{verstraete2008,bridgeman2017,cirac2021} representation that describes a one-parameter family of tRVB states [\cref{fig:intro:state}].
Utilizing exact and approximate TN methods, we show that all the tRVB states considered have $\mathbb{Z}_3$ topological order except for a fine-tuned point, where we establish the presence of a $\mathrm{U}(1) \times \mathrm{U}(1)$ local symmetry that leads to long-range correlations.
We show that topological properties are stable against dilution of the maximally-packed trimer configurations by studying a tensor-network perturbation that encodes all hard trimer configurations with \emph{at most} one trimer on each vertex.
Furthermore, we discuss a general mechanism that explains the enhancement of the local $\mathbb{Z}_3$ conservation law of \cref{fig:intro:config} to a $\mathrm{U}(1) \times \mathrm{U}(1)$ law, yielding long-range correlations and ultimately spoiling $\mathbb{Z}_3$ topological order.
We support our conclusions with further examples of gapless and gapped tRVB states, on the triangular and honeycomb lattices.

Secondly, we consider a diluted trimer model on the square lattice with a single type of trimers, namely the bent blue trimers in \cref{fig:intro:config}.
The corresponding tRVB state [\cref{fig:intro:state} for $t=0$] will be shown to be a gapped $\mathbb{Z}_3$ topological liquid, thus motivating the choice of this geometry.
The model Hamiltonian has a control parameter that tunes the density of trimers.
We compute the ground-state wavefunction via exact diagonalization on periodic clusters and show that a $\mathbb{Z}_3$ topologically ordered phase arises at finite density.
We show that the blockade effect induced by van der Waals interactions in Rydberg atom arrays can be used to engineer the (bent) trimer constraint on the square lattice, by mapping the four possible trimer orientations on a square into four different excited Rydberg atoms.
The effective Rydberg model is equivalent to the diluted trimer model upon neglecting some trimer configurations [cf. \cref{fig:implementation:mapping}].
We prove numerically that removing these configurations from the superposition does not spoil the $\mathbb{Z}_3$ topological nature of the fully-packed tRVB state.
We find that, despite signatures of a topological phase being elusive in the ground state of the Rydberg Hamiltonian, a semi-adiabatic dynamical preparation bolsters the topological character of the prepared state.

The paper is structured as follows.
In \cref{sec:tn} we introduce a one-parameter family of tRVB states on the square lattice, its TN representation, and a one-parameter perturbation that lower the density of trimers, preserving the TN form.
We study the state phase diagram of this two-parameters family and demonstrate the presence of a stable $\mathbb{Z}_3$ topologically ordered phase.
In \cref{sec:lgt} we relate gapped tRVB states to the $\mathbb{Z}_3$ toric code, from which we can define string operators and non-local excitations.
We discuss the condition for which a $\mathrm{U}(1) \times \mathrm{U}(1)$ gauge theory emerges in trimer models, and verify it by considering tRVB states on various lattice geometries.
In \cref{sec:ryd} we introduce the model of diluted bent trimers on the square lattice, analyze its ground-state properties on finite periodic systems, and show that a narrow topological phase emerges.
We outline the implementation of this model in Rydberg atom arrays, point out the differences between the trimer and Rydberg models and discuss the consequences.
Finally, we analyze dynamical preparation protocols to realize $\mathbb{Z}_3$ topologically ordered states in experiments.

\section{tRVB states on the square lattice}
\label{sec:tn}

Trimers on the square lattice can either be bent or straight [cf. blue and red trimers in \cref{fig:intro:config}].
Therefore, we can define a one parameter family of tRVB states on this lattice by weighting each covering $c$ with coefficients that depend on the number of bent and straight trimers [$N_\perp(c)$ and $N_\parallel(c)$ respectively] in the covering.
In particular, we introduce the parameter $\theta \in [0,\pi/2]$ such that the coefficient of a maximally-packed configuration $c$ is
\begin{equation}
    \mathcal{W}(c) = (\cos \theta)^{N_\perp(c)} (\sin \theta)^{N_\parallel(c)}.
    \label{eq:weights}
\end{equation}
The tRVB state then reads

\begin{equation}
    \Ket{ \mathrm{tRVB}(\theta) } = \frac{1}{\mathcal{N}(\theta)} \sum_{\{ c \} } \mathcal{W}(c)\,\Ket{c},
    \label{eq:trvb}
\end{equation}
where $\mathcal{N}(\theta) = \sqrt{\sum_{\{c\}}[\mathcal W(c)]^2}$ is a normalization factor.
The angle $\theta$ changes the relative weight of bent versus straight trimers [see \cref{fig:intro:config}]: in the limit $\theta=0$ ($\theta=\pi/2$) only bent (straight) trimers contribute.

In this section, we analyze the topological properties of this state with tensor network methods.
After studying the one-parameter family in \cref{eq:trvb}, we add another parameter by considering a diluted TN deformation obtained by destroying trimers with a certain probability.
We map out the state phase diagram and demonstrate the stability of the topological phase against dilution.

\subsection{The tensor network representation of the tRVB model}
We turn our attention to the classical statistical-mechanics model whose partition function is the sum of the squared weights in \cref{eq:weights} of all maximally-packed trimer configurations on the square lattice.
This partition function can be interpreted as the norm of the quantum state in \cref{eq:trvb}.
Its tensor-network representation was previously introduced in Ref.~\cite{lee2017}: the partition function $Z$ can be written as the tiling of rank-4 tensors
\begin{equation}
    Z = \includescaledgraphics{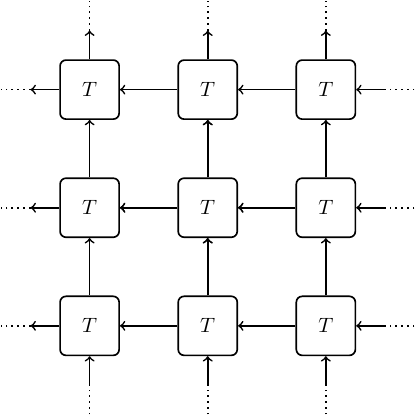}
    \label{eq:partition}
\end{equation}
The rank-4 tensor $T$ is constructed in such a way that, once the tensor is contracted with its neighbors, only the valid trimer configurations survive.
As will be later discussed in \cref{sec:lgt}, it is beneficial to consider an equivalent representation of this model, where trimer configurations are mapped to possible ways to position arrows on the links such that there is a local constraint ---or $\mathbb{Z}_3$ conservation law--- around each edge.
As shown in \cref{fig:intro:config}, a given trimer configuration has a one-to-one mapping to a single arrow configuration obtained by assigning arrows to the links covered by the trimers, in such a way that each arrow goes from the center of the trimer to the external vertices.
In a fully-packed configuration, each vertex has either two outgoing arrows or one ingoing arrow.
Because the net outgoing flux for each vertex is $2$ (where the flux is measured $\mathrm{mod}\,3)$, we obtain that a region of $N_s$ vertices has flux $2N_s\,\mathrm{mod}\,3$.
This $\mathbb{Z}_3$ rule for the flux suggests that the tRVB state can be described as a $\mathbb{Z}_3$ gauge theory and can be a gapped $\mathbb{Z}_3$ quantum spin liquid.

The tensor $T$ is then constructed by labeling each leg with indices $\{0,+1,-1\}$, where $0$ means no arrow, and $+1$ ($-1$) corresponds to an arrow aligned (anti-aligned) with the direction of the leg.
The non-zero entries are
\begin{equation}
    \includescaledgraphics{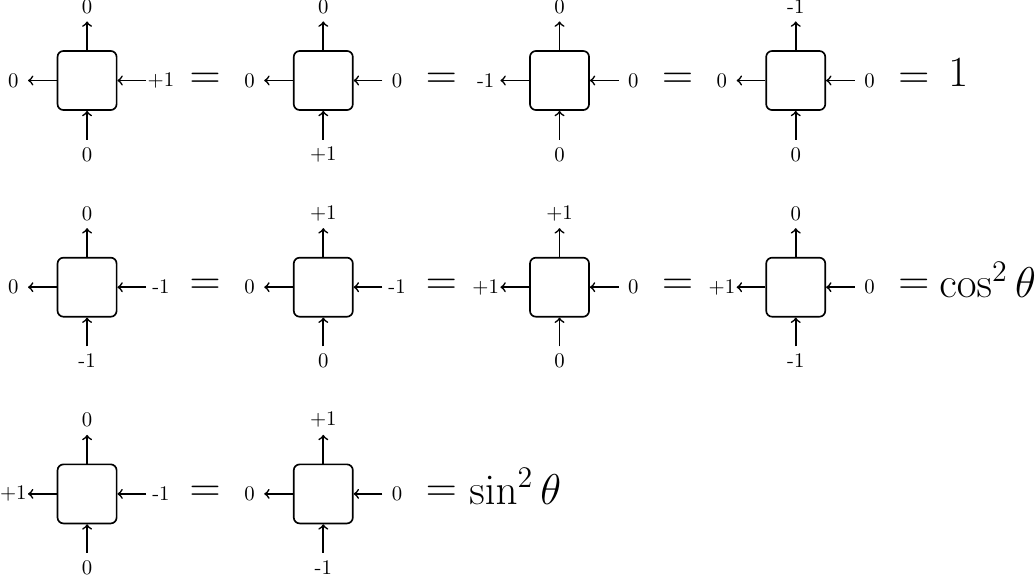} \, .
    \label{eq:tensor_square}
\end{equation}
The $\mathbb{Z}_3$ symmetry of the underlying gauge theory is reflected in the tensor.
Indeed the symmetry operator $\sigma$, whose matrix representation~\footnote{The operator is represented in the basis $\{0,+1,-1\}$ defined above.} is later defined in \cref{eq:sigmatau}, acts on the tensor as
\begin{equation}
    \includescaledgraphics{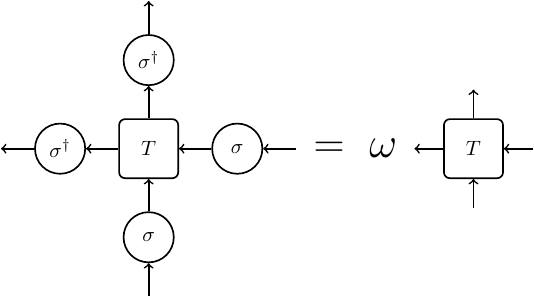}\,.
    \label{eq:symmetry}
\end{equation}
We remark that if one wishes to remove the factor $\omega = e^{i 2 \pi / 3 }$ from \cref{eq:symmetry} it is sufficient to block three consecutive tensors and construct a $\mathbb{Z}_3$-invariant tensor.
When interpreting the partition function encoded in the tensor \cref{eq:tensor_square} as the norm of the tRVB state, each leg of the tensor is interpreted as the product between the bra and ket virtual layers of the quantum state.
\begin{figure}
    \includegraphics[width=\columnwidth]{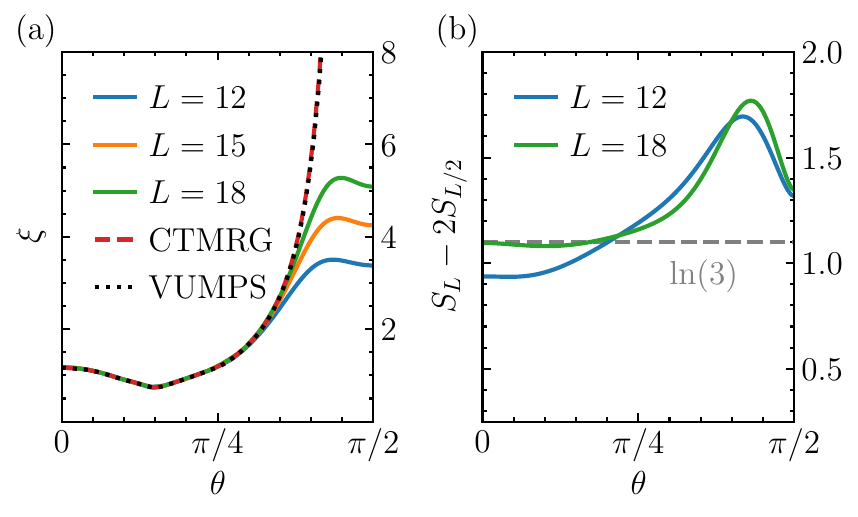}
    \vspace*{-5mm}
    \phantomsubfloat{\label{fig:square:correlation}}
    \phantomsubfloat{\label{fig:square:tee}}
    \caption{%
        (a)~Correlation length $\xi$ as a function of $\theta$, computed using infinite cylinders of circumference $L$, as well as infinite-size methods (see main text).
        The bond dimensions employed for the latter are $D=1536$ and $D=729$ for CTMRG and VUMPS, respectively.
        (b)~Topological entanglement entropy $\gamma \simeq S_L - 2 S_{L/2}$ extracted from finite-size cylinders.}
    \label{fig:square}
\end{figure}
\begin{figure*}
    \includegraphics[width=\textwidth]{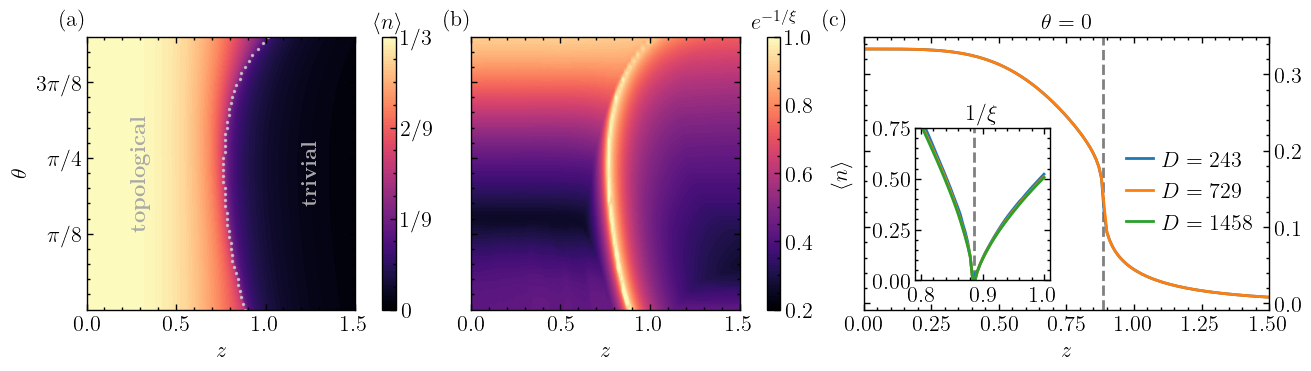}
    \vspace*{-3mm}
    \phantomsubfloat{\label{fig:diluted:density}}
    \phantomsubfloat{\label{fig:diluted:xi}}
    \phantomsubfloat{\label{fig:diluted:detail}}
    \caption{%
        (a)~Density of occupied links of the diluted tRVB state $\ket{\Phi(\theta, z)}$, obtained from CTMRG with an environment bond dimension $D=729$.
        The grey dots indicate where its numerical derivative with respect to~$z$ has an extremum. (b)~Correlation length $\xi$ obtained from the transfer matrix of the diluted tRVB state computed with CTMRG.
        Note that $\xi$ diverges at the phase boundary and at $\theta \to \pi/2$.
        (c)~Details for $\theta=0$: the density displays a non-analyticity at $z_\mathrm{\mathrm{c}} = 0.88(6)$ (dashed line), which corresponds to a divergence of the correlation length (inset).
        The critical point is extrapolated from the maximum $\xi$ at fixed $D$ ($\xi \simeq 50$ for $D=1458$).
    }
    \label{fig:diluted}
\end{figure*}

The properties of the tRVB wavefunction can be extracted by analyzing the row-wise \emph{transfer operator}
\begin{equation}
    \mathbb{T} = \includescaledgraphics{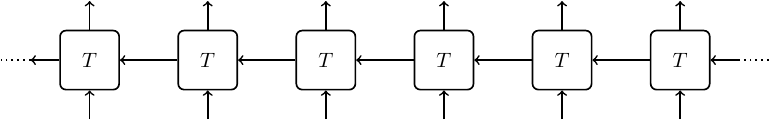}\,.
\end{equation}
By computing the leading eigenvalues (ordered by magnitude) $\lambda_0, \lambda_1, \dots$ of the transfer operator of length $L$ (with periodic boundary conditions), we obtain the correlation length $\xi$ as
\begin{equation}
    \xi = 1/\ln \left|\frac{\lambda_0}{\lambda_1}\right| \, .
\end{equation}
This quantity bounds all correlation functions in the height direction of the infinitely-tall cylinder of circumference $L$.
The $\mathbb{Z}_3$ virtual symmetry in \cref{eq:symmetry} can be used to label the eigenvalues of the transfer operator as $\lambda^{\mathcal{Q}}_k$, where $\mathcal{Q} = 0,\pm 1$ is the $\mathbb{Z}_3$ charge and $k=0,1\dots$ is eigenvalue index starting from the largest in magnitude.
As expected for a topologically ordered state, this symmetry is spontaneously broken, yielding an approximate threefold degeneracy for the largest eigenvalues from the three $\mathbb{Z}_3$ symmetry sectors: the gap between the logarithm of these eigenvalues closes exponentially in $L$~\cite{duivenvoorden2017}.
In fact, the spectrum $E =-\log \lambda$ is analogous to the spectrum of a Hamiltonian with spontaneous symmetry breaking.
To compute the correlation length at finite $L$ we thus consider the two largest eigenvalues in the $\mathcal{Q}=0$ sector.
The results of numerical diagonalizations on finite cylinders are presented in \cref{fig:square:correlation}.
These finite-size results are compared to the correlation lengths obtained from the corner-transfer matrix renormalization group (CTMRG)~\cite{baxter1985,nishino1996,nishino1997,gendiar2001,orus2009,fishman2018} exploiting the reflection symmetry along the tensor diagonal~\cite{chatelain2020}.
For the sake of comparison, in this case we also include results from the variational uniform matrix product state algorithm (VUMPS)~\cite{zauner-stauber2018}.
From the numerical results~\footnote{%
    For finite cylinders calculations we exploited the $\mathbb{Z}_3$ symmetry, blocking three tensors to make the resulting tensor neutral with respect to the symmetry.
    For CTMRG, the halting condition is based on the norm of the difference of the singular values in the corner tensor being below a prescribed tolerance.
    For VUMPS, we use the halting condition of Ref.~\cite{zauner-stauber2018}.
    For both algorithms, we set the tolerance to $10^{-12}$
}, we conclude that the correlation length diverges only in the limit $\theta \to \pi/2$, while it remains finite below that value.
As we will discuss in the next section, long-range correlations concur with the emergence of a $\mathrm{U}(1) \times \mathrm{U}(1)$ local symmetry for $\theta=\pi/2$.
Despite the data do not clearly rule out the presence of an extended interval of diverging $\xi$ in the proximity of $\theta = \pi/2$, the theoretical arguments in the following section, demonstrating that a local continuous symmetry emerges \emph{only} at $\theta = \pi/2$, provide further support to this claim.

The leading eigenvector of the transfer operator on a cylinder encodes the reduced density operator $\rho$ of the infinite half-cylinder~\footnote{%
    The tensor network representation in \cref{eq:tensor_square} is interpreted as the double layer tensor of the tRVB PEPS on the square lattice, where bra and ket are fused in the same virtual leg, and physical indices are summed over.
    One can easily obtain a PEPS by introducing physical qubits $\{ \ket{0} , \ket{1} \}$ on the links, where $\ket{0}$ ($\ket{1}$) stands for the absence (presence) of a trimer on the link.
    Standard tensor network techniques can then be applied (see e.g.~\cite{rdm}) to construct the reduced density operator from the PEPS cylinder transfer matrix.
},
from which we can as well obtain the entanglement entropy $S = -\tr(\rho \ln \rho)$ of this bipartition.
The scaling with the circumference length $L$ of the entanglement entropy obeys
\begin{equation}
    S_L \sim \alpha L - \gamma, \label{eq:tee}
\end{equation}
where $\gamma$ is a well-known topological correction~\cite{kitaev2006,levin2006}.
$\gamma \simeq \ln 3$ implies that the state is in a gapped $\mathbb{Z}_3$ topological phase.
In \cref{fig:square:tee} we plot the topological entanglement entropy obtained from the subtraction $\gamma = S_L - 2 S_{L/2}$, as a function of $\theta$.
While $\gamma$ appears to approach a finite value compatible with $\ln 3$ for $\theta \ne \pi/2$, a bump occurs in the proximity of this point.
The peak position drifts towards $\theta = \pi/2$ as $L$ increases due to finite size effects and we expect it to turn into a singularity at $\theta = \pi/2$ in the thermodynamic limit.
In fact, in the presence of continuous local symmetries such as $\mathrm{U}(1) \times \mathrm{U}(1)$ the topological correction is expected to scale logarithmically with $L$~\cite{dreyer2018,entropy_note}.

\subsection{Stability under dilution of the tRVB state}
We now study a deformation of the tRVB state obtained by diluting fully-packed trimer coverings.
This deformation will be relevant for \cref{sec:ryd}, where we will discuss how to implement trimer models in Rydberg atom arrays.
In these setups, the total occupation can fluctuate, so it is important to consider imperfect trimer coverings.

Similarly to a recent paper on dimer models~\cite{giudici2022}, we consider the following variational ansatz, which depends on two real parameters $0 \leq \theta \leq \pi/2$ and $z \in \mathbb{R}$
\begin{equation}
    \Ket{\Phi (\theta,z)} \propto  \bigotimes_{i,j} \left(\id + z^2 \Sigma^-_{i j}\right) \Ket{\mathrm{tRVB}(\theta)}\, , \label{eq:diluted_trvb}
\end{equation}
where $\Sigma^-_{i j}$ is the operator that removes a trimer on the edges $i$ and $j$, and $z^2$ corresponds to the weight of a removed trimer.
In essence, we add to the fully-packed trimer configurations other trimer configurations that can be obtained from the former by removing trimers without moving the remaining ones.
Each removed trimer is weighted by $z^2$.
In the limit $z \to \infty$, the state is simply the vacuum, while at $z \to 0$ we recover the tRVB state, which we showed to be in a topological phase.
Similarly to $\ket{\mathrm{tRVB}}$, the state $\ket{\Phi}$ has a simple projected entangled-pair state (PEPS)~\cite{cirac2021} representation of bond dimension $4$
\begin{equation}
    \ket{\Phi} = \hspace{-1em}\includescaledgraphics{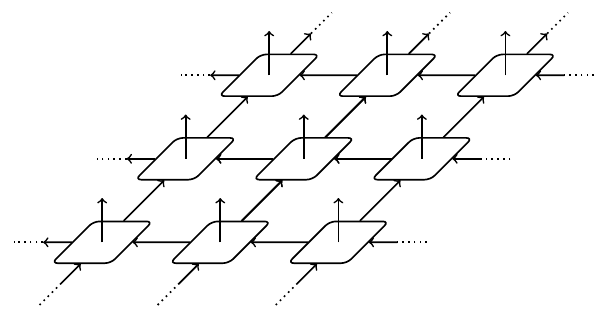}\,.
\end{equation}
Its construction is relegated to \cref{app:details:square}.

In \cref{fig:diluted}, we study the density of occupied links $n$~\footnote{%
    The density of occupied links $n$ counts the ratio of links occupied by a trimer over the total number of links.
    A trimer consists of two occupied neighboring links.
    On the square lattice the density of occupied links of a fully-packed configuration is $1/3$.
}
and the correlation length of the state as we vary $\theta$ and $z$ using CTMRG.
Remarkably, the topological phase survives up to values of order one of the dilution strength $z$, and is fairly insensitive to the mixing angle $\theta$.
The transition between the topological and trivial phases appears to be continuous, as witnessed by a possibly diverging correlation length at the critical point [\cref{fig:diluted:xi}].
It is worth noting that because the bond dimension $D$ of CTMRG environment bond is always finite, we never obtain a truly diverging correlation length (the maximum $\xi \simeq 50$ for $D=1458$).
We do not address the characterization of the universality class of this phase transition as the critical exponents we could extract from the available data exhibit strong dependence on the CTMRG environment bond dimension $D$, even for the largest $D$ we employed.
We note that topological and trivial phases are distinguished by two different degeneracies of the PEPS cylinder transfer matrix.
The spectrum of the transfer matrix is threefold degenerate in the former and ninefold degenerate in the latter.
The ninefold degeneracy reflects the full breaking of the $\mathbb{Z}_3 \times \mathbb{Z}_3$ virtual symmetry of the PEPS double tensor, which implies the condensation of magnetic and the confinement of electric excitations (see \cref{sec:lgt}) in the gauge theory picture~\cite{duivenvoorden2017}.

\section{tRVB states and lattice gauge theories}
\label{sec:lgt}
To understand the emergence of $\mathbb{Z}_3$ topological order in the tRVB state, it is useful to shed light on its connection with a gauge theory.
To this end, in \cref{sec:toriccode} we compare the tRVB state to the ground state of a $\mathbb{Z}_3$ toric code.

\begin{figure}
    \centering
    \includegraphics[width=\linewidth]{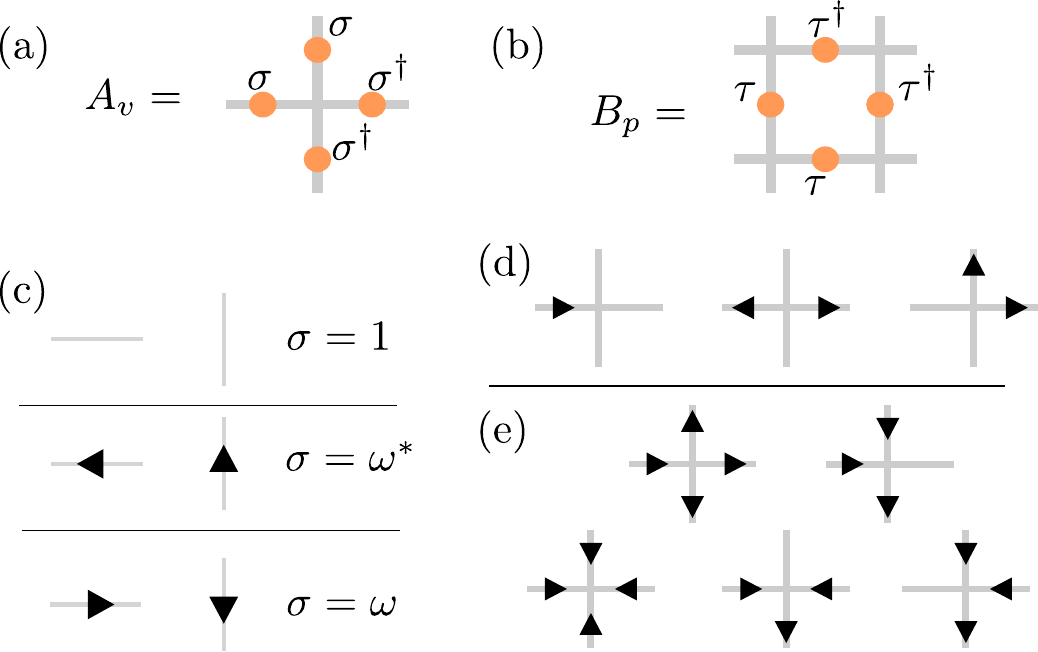}
    \phantomsubfloat{\label{fig:z3gauss:defA}}
    \phantomsubfloat{\label{fig:z3gauss:defB}}
    \phantomsubfloat{\label{fig:z3gauss:mapping}}
    \phantomsubfloat{\label{fig:z3gauss:validconfigs}}
    \phantomsubfloat{\label{fig:z3gauss:invalidconfigs}}
    \caption{%
        Definitions of operators (a)~$A_v$ and (b)~$B_p$.
        (c)~Mapping from a trimer configuration to a state in the $\sigma$ basis. (d)~Configurations for a vertex in the trimer model.
        The 10 total configurations are obtained from these under rotations.
        (e)~Configurations that satisfy Gauss' law but to do not correspond to trimer configurations.
        The full set of unallowed configurations (17 in total) is obtained from these under rotations.
    }
    \label{fig:z3gauss}
\end{figure}

\begin{figure*}
    \centering
    \includegraphics[width=\linewidth]{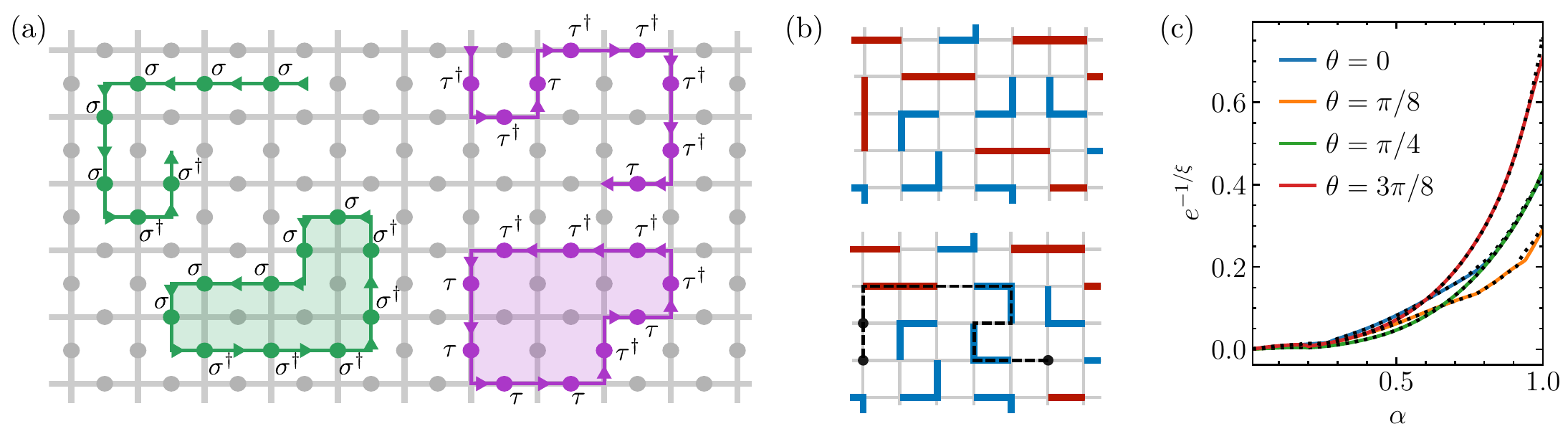}
    \vspace*{-3mm}
    \phantomsubfloat{\label{fig:lines:loops}}
    \phantomsubfloat{\label{fig:lines:excitation}}
    \phantomsubfloat{\label{fig:lines:interpolation}}
    \caption{%
        (a)~String operators in the $\mathbb{Z}_3$ toric code: 't Hooft lines and 't Hooft loops (green), Wilson lines and Wilson loops (purple).
        (b)~Creation of a charge-anticharge pair on a trimer configuration with a string operator: a monomer is an anticharge, and a charge consists of a pair of monomers. (c)~The correlation length remains finite when interpolating between the $\mathbb{Z}_3$ toric code ($\alpha = 0$) and the tRVB state ($\alpha = 1$) at different mixing angles $\theta$ of the $\mathbb{Z}_3$-charged tensor $T = (1 - \alpha) T_\mathrm{TC} + \alpha T_\mathrm{tRVB}(\theta)$.
        In the tensor $T_\mathrm{TC}$, all the components allowed by the virtual symmetry have unit weight (see \cref{fig:z3gauss}).
        Finite-size results on a cylinder of $L=18$ (solid lines) agree with the CTMRG results with $D=729$ (dotted lines).
    }
    \label{fig:lines}
\end{figure*}

We argued in \cref{sec:tn} that the presence of a $\mathbb{Z}_3$ local conservation law of the flux suggests an emergent description as a $\mathbb{Z}_3$ gauge theory, and hence the tRVB state is a good candidate for being a gapped $\mathbb{Z}_3$ quantum spin liquid.
However, as we show below, it may happen that for some trimer models the $\mathbb{Z}_3$ local symmetry is enhanced to a $\mathrm{U}(1)\times\mathrm{U}(1)$ symmetry, in which case the state is gapless~\footnote{As we discuss below, there are exceptions in which, despite the $\mathrm{U}(1) \times \mathrm{U}(1)$ local, the tRVB state is a gapped symmetry broken state.}.
A similar scenario occurs for RVB states of dimer models, that are known to host gapped $\mathbb{Z}_2$ spin liquids only on non-bipartite lattices; on bipartite lattices, they are described by a U$(1)$ gauge theory, that does not support a stable topologically ordered phase.
In \cref{sec:u1u1} we will formulate a similar criterion for trimer models.

\subsection{The \texorpdfstring{$\mathbb{Z}_3$}{ℤ₃} toric code}
\label{sec:toriccode}
A state that is very similar to the tRVB state and has $\mathbb{Z}_3$ topological order is the ground state of the $\mathbb{Z}_3$ generalization of Kitaev's toric code~\cite{kitaev2003}.
We now review this model and show the similarities and differences between its ground state and the tRVB state.

To define the $\mathbb{Z}_3$ toric code, we introduce clock variables on the links of our lattice; on each link we define the operators $\sigma$ and $\tau$, that satisfy the following properties:
\begin{equation}
    \sigma\tau=\omega\tau\sigma,\qquad \sigma^3=\id, \qquad \tau^3=\id,
\end{equation}
where $\omega = e^{2i\pi/3}$.
These variables are the $\mathbb{Z}_3$ generalizations of the Pauli matrices $\sigma^z$ and $\sigma^x$, and their most common matrix representation is
\begin{equation}
    \sigma=
    \begin{pmatrix}
        1 & 0      & 0        \\
        0 & \omega & 0        \\
        0 & 0      & \omega^*
    \end{pmatrix}
    ,\hspace{0.8cm} \tau=
    \begin{pmatrix}
        0 & 0 & 1 \\
        1 & 0 & 0 \\
        0 & 1 & 0
    \end{pmatrix}
    .
    \label{eq:sigmatau}
\end{equation}
We now define the star and plaquette operators as in \cref{fig:z3gauss:defA,fig:z3gauss:defB}.
Similarly to the case of the $\mathbb{Z}_2$ toric code, these operators all commute: $[A_v, B_p]=0$ for every vertex $v$ and plaquette $p$.
We now define the state $\ket{\psi_\text{TC}}$ as the equal-weight superposition of all the states in the $\sigma$ basis that satisfy the Gauss' law $A_v=\omega$ for all vertices.
Note that this choice differs from the typical case with $A_v=1$ and corresponds to the presence of a background charge on each vertex of the lattice.
Nevertheless, the physical properties that we are interested in are not altered by this background charge, as a unitary transformation can be defined to eliminate it.
The state defined here has the property that $B_p \ket{\psi_\text{TC}}= \ket{\psi_\text{TC}}$ for every plaquette $p$, and is the ground state of the following Hamiltonian
\begin{equation}
    H_\text{TC}=-\sum_v (\omega^* A_v+\omega A_v^\dagger)-\sum_p (B_p+B_p^\dagger).
\end{equation}
Because star and plaquette operators commute, it is easy to identify the excitations of the model: we call an excitation with $A_v=\omega^*$ ($A_v=1$) a charge (anticharge), while an excitation with $B_p=\omega$ ($B_p=\omega^*$) is a vison (antivison).
Both ``electric'' (charge/anticharge) and ``magnetic'' (vison/antivison) excitations are gapped.

We now elucidate the connection between the tRVB state and $\ket{\psi_\text{TC}}$.
We can map each configuration of fully-packed trimers to a configuration in the $\sigma$ basis as shown in \cref{fig:z3gauss:mapping}.
It is easily shown that this configuration satisfies Gauss' law.
However, not all the configurations of the $\mathbb{Z}_3$ toric code that satisfy Gauss' law correspond to a trimer configuration: as shown in \cref{fig:z3gauss:validconfigs}, only $10$ of the $27$ configurations of a vertex correspond to allowed vertex configurations of the trimer model.
Despite this difference, the tRVB state may still have $\mathbb{Z}_3$ topological order like $\ket{\psi_\text{TC}}$ if the missing configurations are recovered under renormalization: for $0\le \theta<\pi/2$ the renormalization-group flow from the state $\ket{\text{tRVB}(\theta)}$ ultimately leads to the toric code ground state, which is a fixed point under blocking, with correlation length $\xi=0$.
Another approach to establish that the tRVB state and the toric code ground state describe the same phase consists in showing that no phase transitions occur when interpolating between the two states.
This operation can as well be interpreted as a smooth interpolation between the two parent Hamiltonians, since $\mathbb{Z}_3$ injectivity is preserved~\cite{schuch2010,duivenvoorden2017}.
We interpolate between the states $\ket{\psi_\mathrm{TC}}$ and $\ket{\mathrm{tRVB}}$ by progressively decreasing the weight of the forbidden configurations in \cref{fig:z3gauss:invalidconfigs}.
In \cref{fig:lines:interpolation} we plot the correlation length $\xi$ obtained from CTMRG during the interpolation.
The gradual increase of $\xi$ for any $0\le \theta<\pi/2$ indicates the absence of phase transitions.
We note that what we observed here differs from what happens in the dimer model of the kagome lattice, where the RVB state {\it is} a fixed point of $\mathbb{Z}_2$ topological order, and can be directly mapped into the toric code ground state~\cite{schuch2012}.

Finally, the connection with the $\mathbb{Z}_3$ toric code allows one to define string operators that are useful for detecting topological order, namely Wilson lines and 't Hooft lines.
The latter can be defined as in \cref{fig:lines:loops}.
Because of Gauss' law, the value of the 't Hooft line around a closed loop is equal to $\omega^{N_v+n_q-n_{\bar q}}$, where $N_v$ is the number of vertices, and $n_q$, $n_{\bar{q}}$ are respectively the numbers of charges and anticharges enclosed by the loop.
Similarly, Wilson loops detect the number of visons/antivisons in a region.
Moreover, a 't Hooft (Wilson) line creates a vison/antivison (charge/anticharge) pair at the two ends of the line.

We now consider the same string operators on the trimer model.
The diagonal operator ('t Hooft line) is still well defined.
Closed 't Hooft loops count the number of charges/anticharges in a closed region.
Note that, if we consider a diluted tRVB state, we allow only for the presence of monomers on vertices, i.e., anticharges having $A_v = 1$.
In this case, a pair of monomers represents a charge.
In contrast with the 't Hooft line, the off-diagonal operator (Wilson line) is not well defined on the trimer model, as it can map a valid trimer configuration to one that contains one of the vertices in \cref{fig:z3gauss:invalidconfigs}.
However, as shown in \cref{fig:lines:excitation}, on some states it is possible to define an operator that acts similarly to a Wilson line, and creates a monomer at one end of the line, and a pair or monomers at the other end.
The charges and anticharges obtained in this way are deconfined if the state has topological order.

Knowing the operatorial form of Wilson and 't Hooft lines provides (non-local) order parameters~\cite{bricmont1983,fredenhagen1983} that can be used to assert the presence of $\mathbb{Z}_3$ topological order, as exploited in \cite{verresen2021} for $\mathbb{Z}_2$ topological spin liquids in Rydberg atom arrays.
Although in \cref{sec:ryd} we will not undertake the calculation of these order parameters because of the limited system sizes, we point out that they might be an effective probe for experimental realizations of trimer models.

\subsection{Tripartite trimer models and \texorpdfstring{$\mathrm{U}(1)\times \mathrm{U}(1)$}{U(1)×U(1)} lattice gauge theories}
\label{sec:u1u1}
\begin{figure}
    \centering
    \begin{overpic}[width=\linewidth]{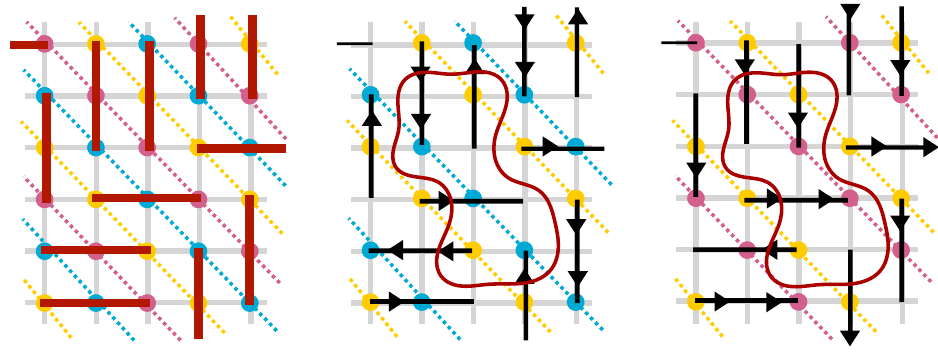}
        \put(0,38){(a)}
        \put(35,38){(b)}
        \put(70,38){(c)}
    \end{overpic}
    \vspace*{-3mm}
    \phantomsubfloat{\label{fig:u1u1:partition}}
    \phantomsubfloat{\label{fig:u1u1:first}}
    \phantomsubfloat{\label{fig:u1u1:second}}
    \caption{%
        (a)~Partition of the square lattice in the three sublattices A (yellow), B (blue), and C (pink).
        A straight trimer always covers one site of type A, one of type B and one of type C.
        (b)~First U$(1)$ symmetry: electric field lines go from the A site to the B site for each trimer.
        The net flux going out of the region enclosed by the red line is $-1=N_A-N_B$.
        (c)~Second U$(1)$ symmetry: electric field lines go from the A site to the C site for each trimer.
        The net flux going out of the region enclosed by the red line is $-1=N_A-N_C$.
    }
    \label{fig:u1u1}
\end{figure}

As shown in \cref{fig:square}, the correlation length of the model diverges for $\theta=\pi/2$, implying that the tRVB state containing only straight trimers is gapless.
We now explain this result, by proving that for straight trimers the $\mathbb{Z}_3$ symmetry is enhanced to a U$(1)\times$U$(1)$ symmetry.
We define a partition of the square lattice in three sublattices A, B, and C as in \cref{fig:u1u1:partition}.
It is easy to check that a straight trimer always covers one and only one site per type.
A similar scenario occurs for dimer models on bipartite lattices: each dimer covers one site of each type, and the symmetry is enhanced from $\mathbb{Z}_2$ to $\mathrm{U}(1)$.
Here, we will show that the emergent symmetry for straight trimers is $\mathrm{U}(1)\times\mathrm{U}(1)$.
To prove it, we define two electric fields.
The first electric field flows from the A to the B site of each trimer [\cref{fig:u1u1:first}], and the second electric field flows from the A to the C site [\cref{fig:u1u1:second}].
Consider a region with $N_A, N_B, N_C$ vertices of the three types:
the net flux going out of the region is $N_A-N_B$ for the first electric field and $N_A-N_C$ for the second electric field.
We then obtain two independent conservation laws, one for each electric field (for a proof of the independence see \cref{app:details:indep}).
Therefore, the tRVB state has a local symmetry $\mathrm{U}(1)\times\mathrm{U}(1)$ and must be gapless, as shown by Polyakov~\cite{polyakov1977}.
This local symmetry can be recasted as a virtual symmetry of the tRVB tensor defined in \cref{eq:tensor_square} (see \cref{app:details:square} for more details).

The argument presented above can be generalized to any trimer model on a two-dimensional lattice.
We posit that a trimer model is tripartite if three sublattices can be defined, such that a trimer always covers one site for each sublattice.
Note that this definition depends both on the lattice and on the class of trimers considered.
If a trimer model is tripartite, the tRVB state has a local U$(1)\times$U$(1)$ symmetry.
In the absence of lattice symmetry breaking, the emergence of this continuous local symmetry leads to a gapless spin liquid state akin to RVB states in dimer models on bipartite lattices.
We can thus conclude that a necessary condition for having a gapped $\mathbb{Z}_3$ spin liquid from a tRVB state is that the trimer model is \emph{not} tripartite.
We remark that this condition is not sufficient, as demonstrated by the examples that we provide below.

\begin{figure}
    \vspace*{2mm}
    \centering
    \includegraphics[width=\linewidth]{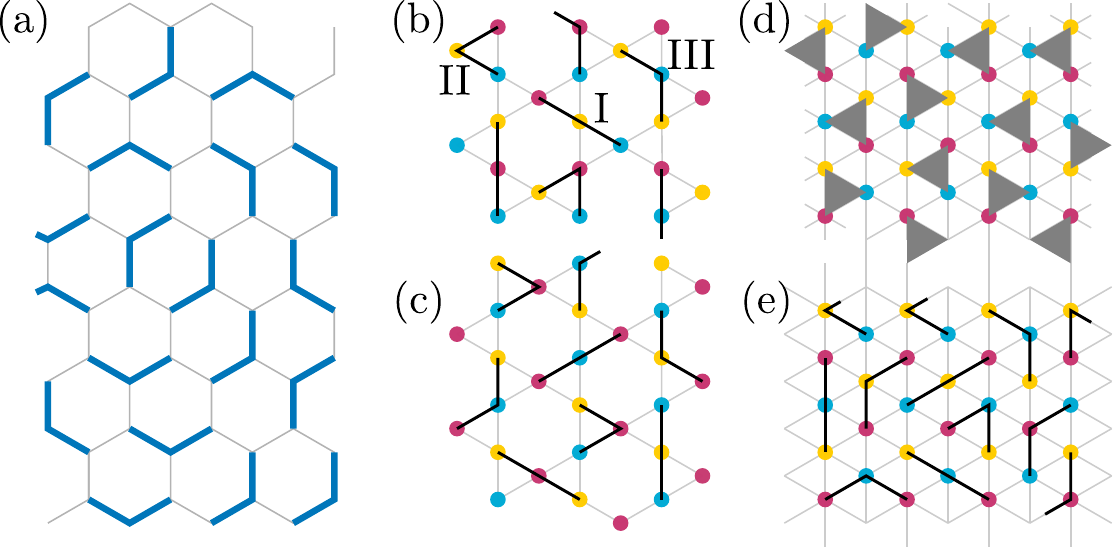}
    \vspace*{-3mm}
    \phantomsubfloat{\label{fig:otherlat:honeycomb}}
    \phantomsubfloat{\label{fig:otherlat:kagome}}
    \phantomsubfloat{\label{fig:otherlat:kagome2}}
    \phantomsubfloat{\label{fig:otherlat:triangular_tripartite}}
    \phantomsubfloat{\label{fig:otherlat:triangular_nontripartite}}
    \caption{(a)~Trimer model on the honeycomb: the lattice is not tripartite. [(b)-(c)] Two possible tripartitions of the trimer model on the kagome lattice.
        We define trimers of type I (straight), II (bent, with angle $60^\circ$), and III (with angle $120^\circ$).
        (b)~The lattice is tripartite if no type III trimers are included. (c)~Similarly, the lattice is tripartite if no type I trimers are included. (d)~Trimer model on the triangular lattice: the lattice is tripartite for triangular trimers. (e)~The triangular lattice is tripartite for trimers of type I and II, not tripartite for trimers of type III.}
    \label{fig:otherlat}
\end{figure}

\begin{figure}
    \phantomsubfloat{\label{fig:otherlat_numerics:honeycomb_tee}}
    \phantomsubfloat{\label{fig:otherlat_numerics:honeycomb}}
    \phantomsubfloat{\label{fig:otherlat_numerics:comparison}}
    \phantomsubfloat{\label{fig:otherlat_numerics:triangular}}
    \includegraphics[width=\columnwidth]{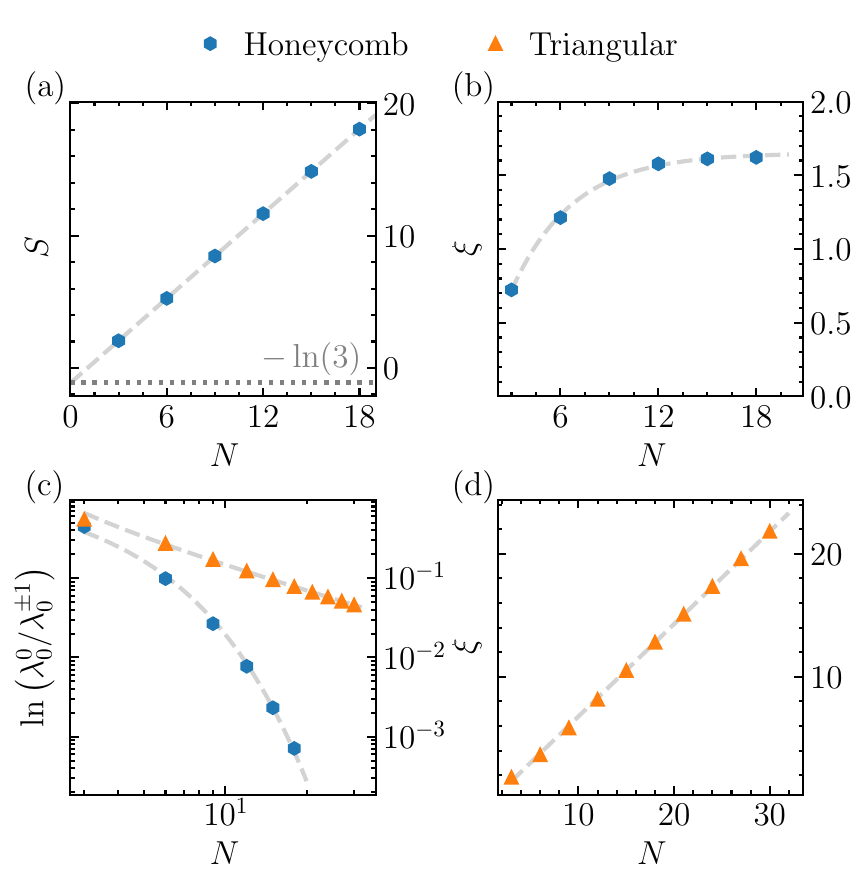}
    \vspace*{-4mm}
    \caption{%
        (a)~Scaling of the entanglement entropy of the tRVB state on the honeycomb lattice as a function of the number of tensors $N$ around the cylinder.
        We extract a topological correction $\gamma \simeq \ln 3$.
        (b)~Correlation length of the tRVB state on the honeycomb lattice. (c)~Logarithmic gap between the leading eigenvalues of the $\mathcal{Q}=0$ and $\mathcal{Q}=\pm 1$ sectors for the tRVB state on the honeycomb (blue hexagons) and triangular (orange triangles).
        On the triangular lattice, only triangular trimers are considered.
        Transfer matrix eigenvalues are labeled as $\lambda^{\mathcal{Q}}_n$ where $\mathcal{Q}$ is the $\mathbb{Z}_3$ symmetry sector and $n=0,1\dots$ is the position in the spectrum starting from the largest in magnitude.
        (d)~Correlation length of the tRVB on the triangular lattice.
        Similar to dimer models on bipartite lattices, it grows linearly with the number of tensors $N$ along the circumference.
        The TN description of these models can be found in \cref{app:details}.} \label{fig:otherlat_numerics}
\end{figure}

Let us first consider the tRVB state on the honeycomb lattice [\cref{fig:otherlat:honeycomb}].
This trimer model is not tripartite, so this state on the honeycomb lattice can be a gapped state with $\mathbb{Z}_3$ topological order.
The numerics in \cref{fig:otherlat_numerics} confirm that this is the case.
In \cref{fig:otherlat_numerics:honeycomb} we show that the correlation length of the tRVB state on a finite cylinder converges to a finite value as the circumference increases.
In \cref{fig:otherlat_numerics:honeycomb} we show that the entanglement entropy of a half-infinite cylinder exhibits a $-\ln 3$ correction to its area law scaling.
Finally, the blue hexagons in  \cref{fig:otherlat_numerics:comparison} demonstrate that the logarithmic gap between the leading eigenvalues of the neutral and charged sectors closes exponentially, pointing to the spontaneous breaking of $\mathbb{Z}_3$ virtual symmetry.

Let us now turn our attention to the kagome lattice.
In this case, various types of trimers can be defined.
If we consider the tripartition of the lattice shown in \cref{fig:otherlat:kagome}, we note that some types of trimers (I and II) cover sites of different types, while trimers of type III do not.
Therefore, we deduce that type III trimers are needed to have a $\mathbb{Z}_3$ spin liquid phase.
Similarly, from the tripartition in \cref{fig:otherlat:kagome2}, we find that type II trimers are also needed.
This result is in agreement with Ref.~\cite{jandura2020}, where it was shown that a gapped tRVB state with topological order is found only when all types of trimers are included.
This lattice provides a counterexample that shows how our ``non-tripartibility'' condition is not sufficient for gapped $\mathbb{Z}_3$ topological order.
In fact, in Ref.~\cite{jandura2020} it was proven that the tRVB state with trimers of type I and III possesses a $\mathrm{U}(1)$ local symmetry that spoils $\mathbb{Z}_3$ topological order although the trimer model cannot be tripartite.
Trimer models on this geometry also provide an example of $\mathrm{U}(1) \times \mathrm{U}(1)$ symmetric tRVB state that is not gapless but has symmetry breaking: the tRVB state with trimers of type II only is tripartite, but {\it all} trimer coverings break the two-fold rotation of the lattice that maps upper into lower triangles.

Finally, let us consider the triangular lattice.
As can be inferred from \cref{fig:otherlat:triangular_tripartite}, the tRVB state of triangular trimers~\footnote{
    The triangular trimers considered here do not consist of two edges sharing a common vertex, but rather of triangular plaquettes with hard constraints.
    Nevertheless, the discussion on the general conditions for $\mathbb{Z}_3$ or U$(1)\times$U$(1)$ gauge symmetry is applicable here in the same way.
} (grey triangles in the figure) is $\mathrm{U}(1) \times \mathrm{U}(1)$ symmetric because the model is tripartite: this finding agrees with Ref.~\cite{verberkmoes2001}, where a U$(1)\times$U$(1)$ conservation law (for ``left-'' and ``right-movers'') was found in the classical configurations.
From numerical diagonalization of the transfer matrix on finite-size cylinders, we deduce that this tRVB state is indeed gapless, as demonstrated in \cref{fig:otherlat_numerics:triangular}, where we show that for the sizes accessible with our numerics, the correlation length scales linearly with the circumference of the cylinder.
We refer to \cref{app:details:triangular} for the explicit TN representation of this tRVB state.
Using the same definitions of trimers as for the kagome lattice, we have that trimers of type I (straight) on the triangular lattice are also tripartite and expected to spoil gapped topological order, while trimers of type III do not respect the tripartition [\cref{fig:otherlat:triangular_nontripartite}].
This implies that tRVB states on this lattice can have topological order only if these trimers are included.
We leave a complete analysis of this family of tRVB states for future work.

\begin{figure}
    \includegraphics[width=\columnwidth]{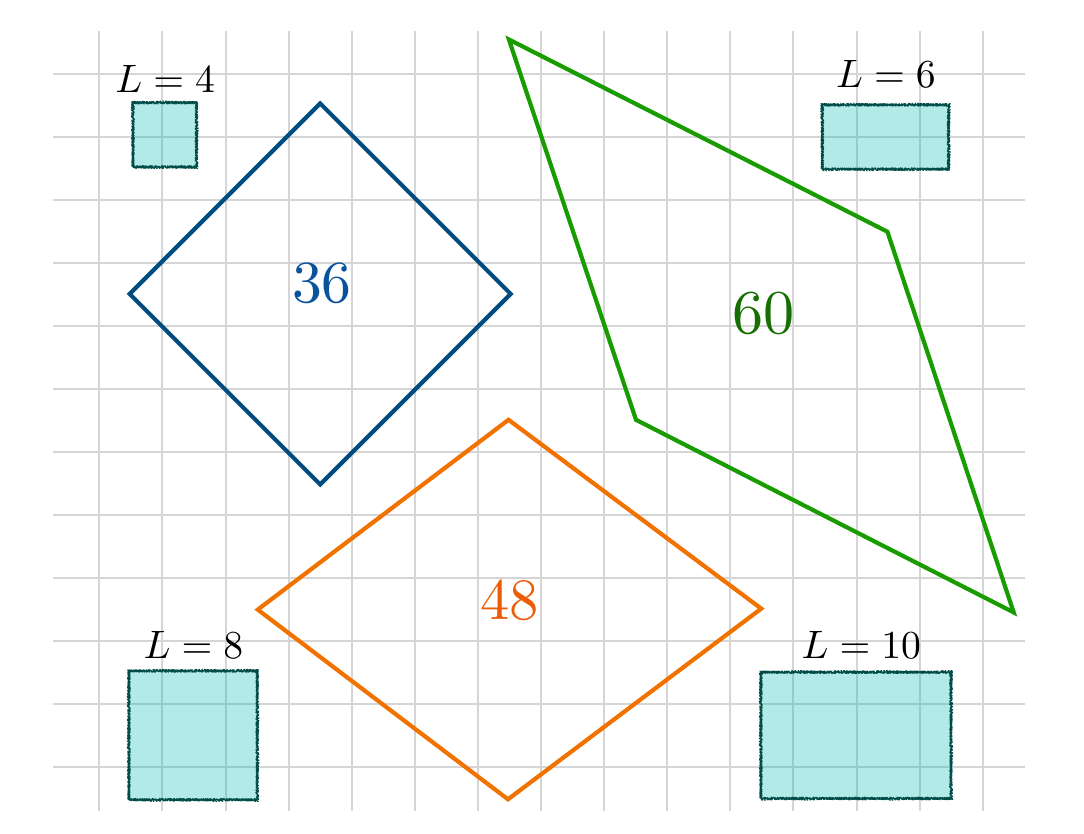}
    \caption{%
        Periodic clusters employed for the exact diagonalization of the Hamiltonian in \cref{eq:trim_ham}.
        $36$, $48$, and $60$ stand for the number of edges inside the cluster.
        These sizes are chosen such that the number of links is a multiple of $6$, to avoid imperfect trimer coverings at large density.
        The turquoise regions are the subsystems used to compute the entanglement entropy.
        We measure their perimeter $L$ in units of an edge of the square lattice.
    }
    \label{fig:ryd_cluster}
\end{figure}

\section{Diluted trimer models and Rydberg atoms}
In the previous sections we have shown that RVB states of trimers can be gapped and have topological character.
When these conditions are met they are good candidates for representing a stable phase with $\mathbb{Z}_3$ topological order.
It is thus natural to ask if simple Hamiltonians exist that have tRVB-like phases at zero temperature.
As trimer states are TN states with finite bond dimension, they are exact ground states of local Hamiltonians with finite range.
However, it is known that such Hamiltonians can be rather complex and include fairly unphysical operators~\cite{schuch2010,schuch2012,giudici2022a}.
In particular, parent Hamiltonians of tRVB states on certain lattices are discussed in Refs.~\cite{lee2017,jandura2020}.
Here, instead, we introduce a simple trimer model on the square lattice and study its ground state phase diagram via exact methods, providing evidence of a tRVB-like phase with $\mathbb{Z}_3$ topological order.
Moreover, we show that a similar model can be implemented in Rydberg atom arrays and that hallmarks of $\mathbb{Z}_3$ topological order can be observed employing semi-adiabatic dynamical preparation schemes.

\begin{figure}
    \includegraphics[width=\columnwidth]{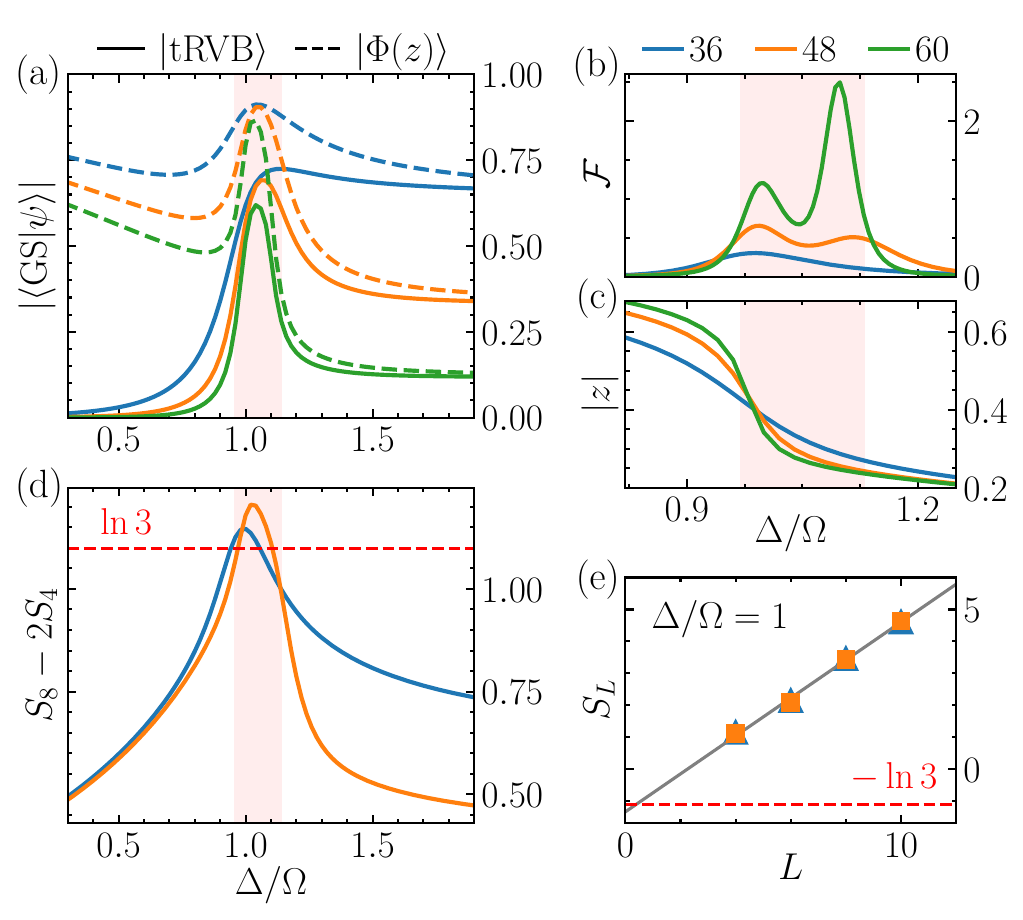}
    \vspace*{-3mm}
    \phantomsubfloat{\label{fig:ryd_gs:overlap}}
    \phantomsubfloat{\label{fig:ryd_gs:fidelity}}
    \phantomsubfloat{\label{fig:ryd_gs:z}}
    \phantomsubfloat{\label{fig:ryd_gs:tee}}
    \phantomsubfloat{\label{fig:ryd_gs:tee2}}
    \caption{%
        (a)~Overlap between the ground state of the Hamiltonian \cref{eq:trim_ham} and the tRVB state (solid line), and the diluted tRVB state \cref{eq:diluted_trvb} optimized over $z$ (dashed line).
        The shaded red region set approximate boundaries for the topologically ordered phase, where the fidelity with the tRVB state is maximized.
        (b)~Ground state fidelity susceptibility per link $\mathcal{F} = (1 - \left|\braket{\mathrm{GS}(\lambda) | \mathrm{GS}(\lambda+ d \lambda)}\right|)/ N d \lambda^2$, with $\lambda = \Omega/\Delta$.
        Two peaks appear for the larger clusters, pointing at the presence of an intermediate phase for $ 0.95 \lesssim \Delta/\Omega \lesssim 1.15 $.
        (c)~Optimal value of $z$ that maximizes the overlap between the ground state and the diluted tRVB state \cref{eq:diluted_trvb}. $|z| \lesssim 0.4$ in the intermediate phase, a value that lies deep in the topological phase in the state phase diagram plotted in \cref{fig:diluted}.
        (d)~Ground state topological entanglement entropy computed by subtracting the entropies of the square-shaped regions in \cref{fig:ryd_cluster} with $L=4$ and $L=8$. (e)~Scaling of the entanglement entropy of the ground state for $L=4,6,8,10$ and $\Omega/\Delta = 1$.
    }
    \label{fig:ryd_gs}
\end{figure}

\subsection{An effective trimer models on the square lattice}
\label{sec:ryd}
We consider the Hilbert space spanned by all diluted trimer configurations of bent trimers on the square lattice, i.e. with at most one trimer per vertex of the square lattice, and take the following model Hamiltonian
\begin{equation}
    H = \frac{\Omega}{2} \sum_{\square} \Ket{\includescaledgraphics{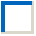}}\!\Bra{\includescaledgraphics{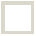}} + \mathrm{H.c.} - \Delta \Ket{\includescaledgraphics{trimer}}\!\Bra{\includescaledgraphics{trimer}} + R_{\frac{\pi}{2}}.
    \label{eq:trim_ham}
\end{equation}
$R_{\frac{\pi}{2}}$ denotes the terms that can be obtained by 90-degree rotations from those given in \cref{eq:trim_ham}.
The first term coherently creates and destroys bent trimers (subject to the hard trimer constraint), whereas the second, diagonal term acts like a chemical potential for trimers.
The ratio $\Delta/\Omega$ controls the density of trimers in the ground state.
For large and negative $\Delta/\Omega$ the ground state is trivial and adiabatically connected to the vacuum.
For $\Delta/\Omega=+\infty$ the classical ground space is exponentially degenerate and consists of all maximally-packed trimer coverings, corresponding to a link density $\braket{n} = 1/3$.
By treating pertubatively the off-diagonal diagonal term, it is easy to see that the first non trivial process in this subspace occurs at fourth order and produces resonances between pairs of trimer coverings differing only on two nearby squares.
Therefore, at large $\Delta/\Omega$ a valence bond solid (VBS) ground state is expected to emerge, with a maximal density of resonating ``plaquettes'', i.e., resonating pairs of nearby squares~\footnote{We note that the number of plaquette coverings is not finite when the system is infinite, and a quantum order by disorder scenario is likely to occur.}.
At finite $\Delta/\Omega$, quantum fluctuations act in two ways: they create defects in the trimer coverings by lowering the density and build coherent superpositions of high density components.
As we showed in \cref{sec:tn}, topological order can survive at finite dilution, implying that a diluted tRVB state might also arise from this setup.

To understand the character of the ground state at intermediate $\Delta/\Omega$ we performed exact diagonalization calculations on periodic clusters of up to $60$ edges of the square lattice [\cref{fig:ryd_cluster}].
In \cref{fig:ryd_gs:overlap} we plot the overlap between the ground state and the pure tRVB state (solid line), and the ground state fidelity with the diluted tRVB state \cref{eq:diluted_trvb} for $\theta = 0$ optimized over $z$ (dashed line).
The optimal values of $z$ as a function of $\Delta/\Omega$ are shown in \cref{fig:ryd_gs:z}.
Remarkably the overlap displays a maximum at $\Delta/\Omega \simeq 1$, pointing at the presence of an intermediate tRVB-like phase.
The maximum fidelity is greatly improved when optimized with the diluted tRVB state.
We note that the optimal value of $z$ near the maximum is perfectly consistent with the topologically ordered phase in the state phase diagram in \cref{fig:diluted}.
The occurrence of an intermediate phase is also witnessed by the presence of two peaks in the ground-state fidelity susceptibility per link $\mathcal{F} = (1 - |\braket{\mathrm{GS}(\lambda)|\mathrm{GS}(\lambda+ d \lambda)}|)/ N d \lambda^2$, where $\lambda = \Omega/\Delta$~\footnote{The parameter $\lambda$ is chosen to make more distinguishable the two peaks in $\mathcal{F}$.} for the $48$- and $60$-links clusters, as depicted in the \cref{fig:ryd_gs:fidelity}.
To confirm the nature of the intermediate phase in an unbiased way, in \cref{fig:ryd_gs:tee} we show the topological entanglement entropy extracted from $\gamma \simeq S_{2 L} - 2 S_L $ as functions of $\Delta/\Omega$.
Here $L$ is the length of the contour of the subsystem, in units of one edge of the square lattice, and the subsystems employed for the computation are depicted in \cref{fig:ryd_cluster}.
The value of $\gamma$ obtained near the tRVB fidelity maximum is remarkably close to the value $\ln 3$, hinting at the emergence of $\mathbb{Z}_3$ topological order.

\subsection{The trimer constraint with Rydberg atoms}
\begin{figure}
    \includegraphics[width=\columnwidth]{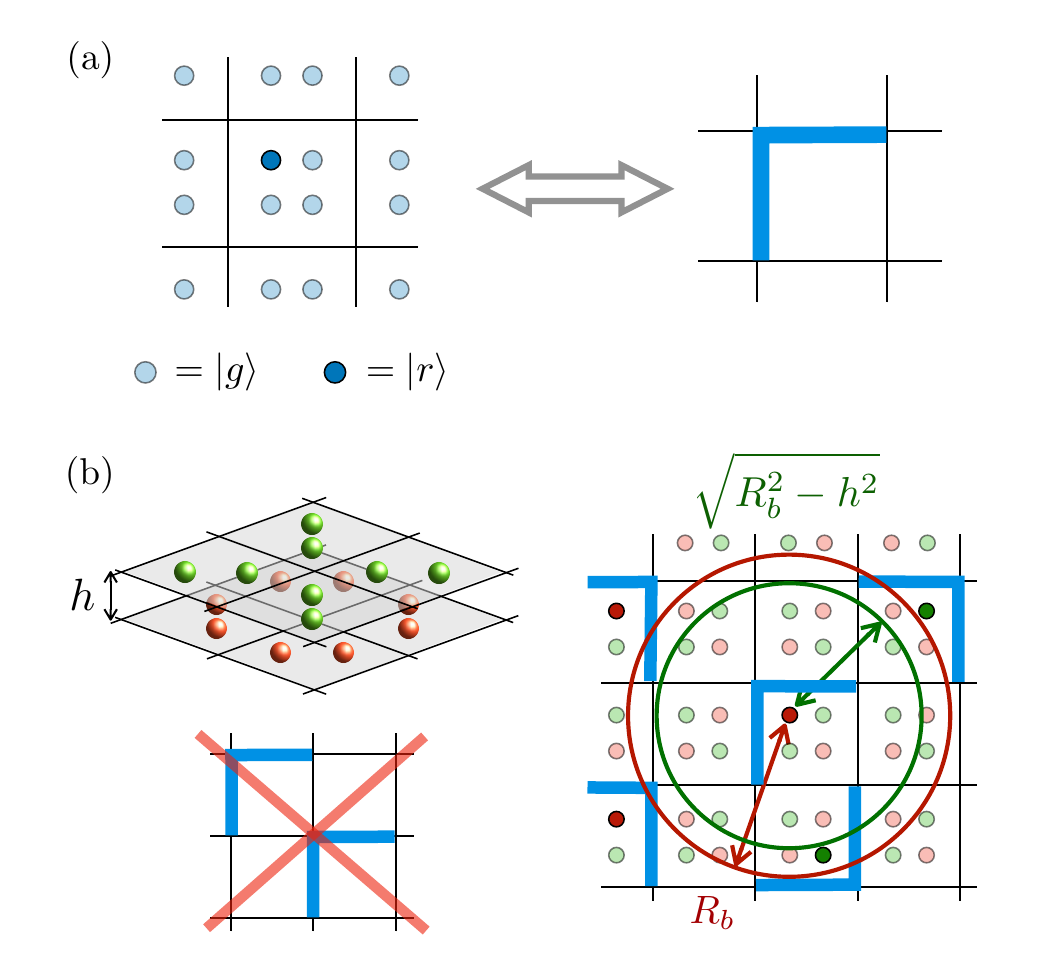}
    \vspace*{-3mm}
    \phantomsubfloat{\label{fig:implementation:mapping}}
    \phantomsubfloat{\label{fig:implementation:lattice}}
    \caption{%
        (a)~Mapping between Rydberg atoms and bent trimers on the square lattice.
        Atoms are placed at each corner of the square lattice such that an excited atom is mapped to a bent trimer on that corner.
        (b)~Atoms are split into two groups (green and red) depending on the sublattice of the square lattice bipartition to which the corresponding corner belongs.
        The groups are arranged onto two planes at distance $h$ such that the 2D blockade radius between atoms of different colors is $\sqrt{R_b^2-h^2}$, where $R_b$ is the 3D blockade radius.
        Tuning $h$ and the distance of the atoms from the vertexes allows realizing a constraint equivalent to the trimer constraint where the ``wedged" trimer configurations on the bottom left are blockaded (and its 90-degree rotations).
    }
    \label{fig:implementation}
\end{figure}

We now turn to a discussion of potential realizations of trimer models and tRVB states with experiments based on Rydberg atom arrays.
In these systems, neutral atoms are individually trapped and arranged in a desired lattice configuration using optical tweezers~\cite{endres2016,barredo2016}.
Spin models can then be realized by manipulating the internal degrees of freedom of each atom with an external laser field~\cite{weimer2010,labuhn2016,bernien2017,deleseleuc2018,kim2018,schauss2015}.
Specifically, we consider a situation where a laser induces a coherent coupling from the atomic ground state $\ket{g}$ to a highly excited Rydberg state $\Ket{r}$.
The frequency mismatch between the laser frequency and the transition frequency between those two states, i.e. the laser detuning, is denoted by $\Delta$.
The coupling strength for this transition, i.e. the Rabi frequency, is denoted by $\Omega$, and is proportional to the laser amplitude.
Importantly, two atoms that are both in the Rydberg state interact via a van der Waals process, whose strength decays with the sixth power of the atomic separation.
As a result, the Hamiltonian governing the dynamics of this system is given by~\cite{browaeys2020}
\begin{equation}
    \label{eq:ryd_ham} H_{\mathrm{Ryd}} = \frac{\Omega}{2} \sum_i \sigma_i^x - \Delta \sum_i n_i +  C \sum_{i>j} \frac{n_i n_j}{|\vec{x}_i-\vec{x}_j|^6} \quad ,
\end{equation}
where, $\vec{x}_i$ is the position of atom $i$, and we defined $\sigma_i^x = \Ket{g}_i\!\Bra{r} + \Ket{r}_i\!\Bra{g} $ and $n_i = \Ket{r}_i\!\Bra{r}$.
The parameter $C$ depends on the Rydberg state.
The interplay between the laser parameters and the geometry of the atom arrangement gives rise to a variety of phenomena~\cite{turner2018,surace2020,semeghini2021,ebadi2021,scholl2021,ebadi2022}.
Most of them are based on the Rydberg blockade effect, that prohibits the simultaneous excitation of two atoms located at a distance $r < R_b = (V/\Omega)^{1/6}$.
Below we show that this effect can be used to implement trimer constraints in Rydberg atom arrays.
For example, it is easy to prove that the hard trimer constraint for triangular trimers on the triangular lattice in \cref{fig:otherlat:triangular_tripartite} is equivalent to a Rydberg blockade constraint on a honeycomb lattice: the atoms sit on the centers of the original triangular lattice and a Rydberg excitation represents a triangular plaquette; the blockade radius $R_b$ is chosen such that two atoms cannot be simultaneously excited if and only if they belong to the same hexagon.
We now show that the Rydberg blockade effect also allows one to realize a bent trimer model on the square lattice very similar to the one outlined above.
A sketch of the implementation is depicted in \cref{fig:implementation}.
Rydberg atoms are placed on the corners of a square lattice such that an excited atom is mapped to a bent trimer.
The basic idea is to exploit the blockade radius to mimic the hard trimer constraint.
However, while the latter is anisotropic, the blockade effect is not, as long as the Rydberg state is rotational invariant.
Nevertheless, we can avoid the use of anisotropic Rydberg states by dividing the atoms into two groups, according to the sublattice of the square to which they are closer.
The two groups are then arranged onto two planes at a distance $h$.
Consequently, atoms between different planes will be blockaded if their planar distance is less than $\sqrt{R_b^2-h^2}$, where $R_b$ is the blockade radius.
By properly choosing $h$ and the atoms positions inside the plaquettes of the square lattice, it is possible to realize a trimer constraint, as demonstrated by \cref{fig:implementation:lattice}.
This trimer constraint is such that some trimer configurations are locally prohibited.
The latter are trimer coverings that include two trimers with the same orientation that are ``wedged'' diagonally as in \cref{fig:implementation:lattice}.

\begin{figure}
    \includegraphics[width=\columnwidth]{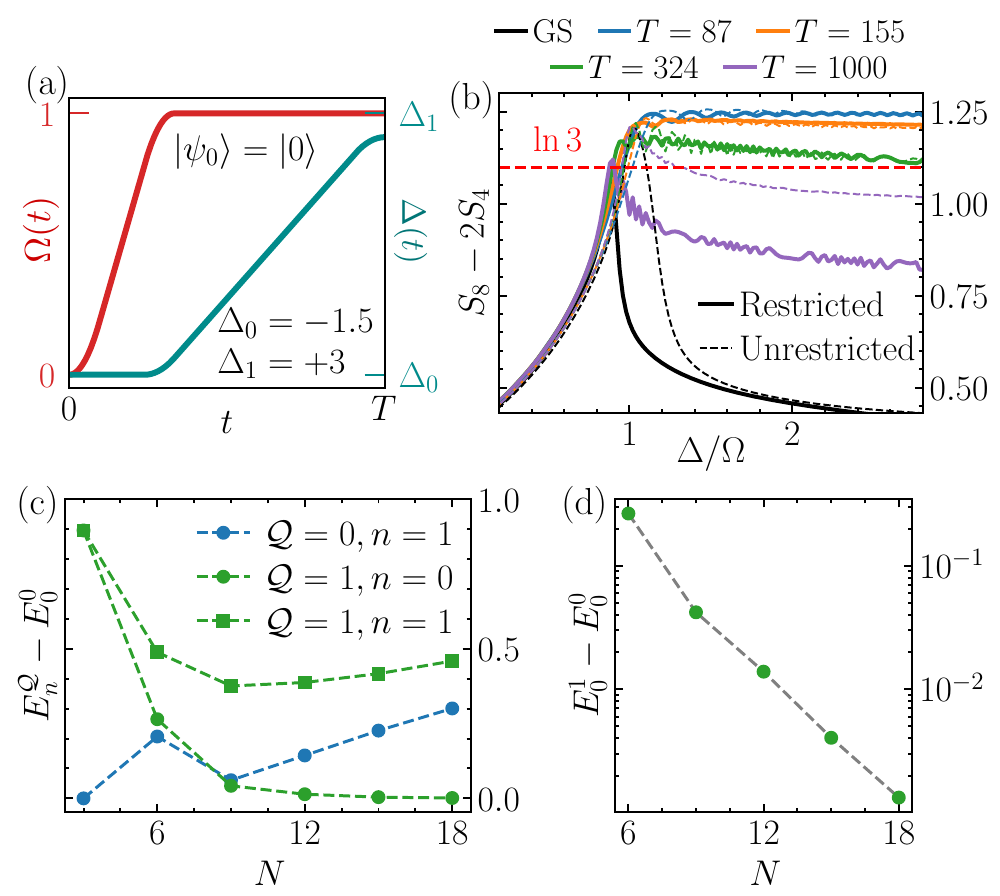}
    \vspace*{-3mm}
    \phantomsubfloat{\label{fig:ryd_dyn:protocol}}
    \phantomsubfloat{\label{fig:ryd_dyn:tee}}
    \phantomsubfloat{\label{fig:ryd_dyn:tm1}}
    \phantomsubfloat{\label{fig:ryd_dyn:tm2}}
    \caption{%
        (a)~The protocol employed for the semi-adiabatic dynamical preparation.
        The vacuum $\Ket{0}$ is evolved with a time-dependent Hamiltonian with $\Delta(t)$ and $\Omega(t)$ as depicted in the figure.
        First, $\Omega$ is switched on from $0$ to $1$ at constant $\Delta = \Delta_0$ with a smoothed linear ramp.
        $\Delta$ is then increased linearly from $\Delta_0=-1.5$ to $\Delta_1=+3$ of constant $\Omega$.
        The total sweep duration is $T$.
        (b)~Topological entanglement entropy of the dynamically prepared state, for different preparation times.
        The black line corresponds to the ground state of the Hamiltonian \cref{eq:trim_ham} ($T=\infty$).
        Solid and dashed lines are obtained in the effective Rydberg model where the ``wedged'' trimer configuration of \cref{fig:implementation:lattice} are not included in the Hilbert space and the unrestricted diluted trimer model, respectively.
        Results are obtained on the periodic cluster $48$ of \cref{fig:ryd_cluster}.
        (c)~$E=-\ln \lambda$, where $\lambda$ are the eigenvalues of the cylinder transfer matrix of the tensor network representation of the restricted tRVB state of bent trimers on the square lattice. $N$ is the number of tensors along the circumference, $\mathcal{Q}$ is the $\mathbb{Z}_3$ virtual charge ($\mathcal{Q}=\pm 1$ sectors are exactly degenerate), $n$ is the eigenvalue index in the sector with charge $\mathcal{Q}$. (d)~Exponential scaling of the gap between the smallest $E$s in the neutral and charged sectors, signaling spontaneous symmetry breaking of the $\mathbb{Z}_3$ virtual symmetry of the tensor. }
    \label{fig:ryd_dyn}
\end{figure}

Before addressing the Rydberg model, we study the effect of removing these coverings from the fully-packed tRVB state.
In \cref{fig:ryd_dyn:tm1,fig:ryd_dyn:tm2} we plot the lowest logarithmic gaps in the spectrum of the cylinder transfer matrix of the corresponding TN state as functions of the cylinder circumference $L$.
The TN representation is outlined in \cref{app:details:square}.
Despite a level crossing occuring at finite $L$, the gap between the neutral and charged sectors eventually closes exponentially (green circles), whereas the neutral gap (blue circles) appears to be increasing for the available $L$s.
From infinite-size calculations we can infer that this gap converges to $\simeq 0.6$, yielding a correlation length $\xi \simeq 1.7$.
We note that this value is larger than the correlation length of the unrestricted trimer state $\xi \simeq 1.1$ [cf. \cref{fig:square:correlation}].
This fact is expected, as removing these configurations pushes away the tRVB state from the $\mathbb{Z}_3$ toric code fixed point, for which $\xi=0$.
These results demonstrate that $\mathbb{Z}_3$ topological order is preserved.
Although we did not study TN perturbations that lower the density of trimers, we expect a diluted version of this tRVB state to host a topologically ordered phase.

We now focus on the Rydberg model arising from the implementation explained above.
For simplicity, we neglect interactions beyond the blockade, so that the effective Hamiltonian is the same as \cref{eq:trim_ham}, with the caveat that all the states containing wedged trimers as in \cref{fig:implementation:lattice} are not included in the Hilbert space of diluted trimer coverings.
The exact diagonalization of this restricted trimer model displays no evidence of an intermediate topological phase in the ground state, rather a single phase transition between a disordered phase and a plaquette phase can be identified.
Therefore, we conclude that if such a phase exists it is extremely narrow.
In fact, the topological entanglement entropy extracted from the finite size cluster $48$ in \cref{fig:ryd_cluster} exhibits a peak approaching $\gamma = \ln 3$ that is much sharper than in the unrestricted model, as we show in \cref{fig:ryd_dyn:tee}.
The black lines are the ground-state curves for $\gamma$ in the restricted (solid line) and unrestricted (dashed line) trimer models.
However, below we provide numerical evidence that this witness of $\mathbb{Z}_3$ topological order is stabilized by a dynamical preparation protocol regularly used in experiments~\cite{semeghini2021}.
\\
The initial state is the vacuum, subsequently evolved with the time-dependent Hamiltonian $H(t) = H (\Omega(t),\Delta(t))$.
To prepare the ground state of $H(t)$ the variation of the time-dependent couplings has to be perfectly adiabatic.
In real experiments, this is very hard in practice, due to limited coherence time.
Thus it is often preferable to consider non adiabatic state preparation schemes.
In fact, as demonstrated in Refs.~\cite{semeghini2021,giudici2022}, non adiabatic effects can even enhance topological order in the prepared state with respect to the ground state.
In the following, we show that a similar result is observed here.
Specifically, we study the dynamical preparation process depicted in \cref{fig:ryd_dyn:protocol} and described below. The latter is the simplest possible preparation protocol to drive the system from the vacuum to a state with a high density of Rydberg excitations.
The vacuum state $\Ket{0}$ is evolved with the time-dependent Hamiltonian $H (\Omega(t),\Delta(t))$, where $\Delta(0) = -2.5$, $\Omega(t)=0$, such that $\Ket{0}$ is the ground state at $t=0$.
A first (smoothed) linear ramp turns on the effective Rabi frequency until the final value $\Omega=1$ is reached.
The latter sets our units of energy and time.
A second ramp is used to drive the detuning from $\Delta(0)=-1.5$ to $\Delta(T)=3$, where $T$ is the total sweep time, and the final value of $\Delta$ is chosen to be well beyond the peaks in tRVB overlap and topological entanglement entropy of the ground state.
The slopes of the two ramps decrease with increasing $T$ and are fixed by requiring that the switching on of $\Omega$ ($\Delta$) takes $T/3$ ($2 T/3$).
In both the restricted and unrestricted models a phase transition is crossed during the second ramp.

In \cref{fig:ryd_dyn} we plot the topological entanglement entropy of the state during the preparation sweep, for different total sweep times $T$, for the $48$ cluster in \cref{fig:ryd_cluster}, that corresponds to $96$ atoms in the mapping of \cref{fig:implementation:mapping}.
The result indicates that topological properties are stabilized in the prepared state when the preparation is {\it not} adiabatic, i.e., for short and intermediate $T$.
Remarkably, the peaks in the topological entropy correction disappear in this regime, and the latter points to a topological state when $\Delta/\Omega \gtrsim 1$.
For the largest $T$s the ground state curve is recovered ($T=\infty$).
We note that this phenomenon occurs in both the restricted and unrestricted models [cf. dash and solid lines in \cref{fig:ryd_dyn:tee}].
We remark that it might be possible to engineer other implementations of the Hamiltonian in \cref{eq:trim_ham} that do not require a restriction of the diluted trimer Hilbert space.

\section{Conclusions}
We showed that maximally-packed trimer states can be simple representatives of quantum spin liquids with $\mathbb{Z}_3$ topological order.
By mapping trimer configurations into the Hilbert space of a lattice gauge theory, we identified a condition on the lattice geometry and trimer model that leads to the emergence of a $\mathrm{U}(1) \times \mathrm{U}(1)$ symmetry and a tRVB state with infinite correlation length.
We verified this condition by performing numerical checks on several trimer models with TN methods.
We demonstrated that when tRVB states are gapped, $\mathbb{Z}_3$ topological order is stable against fluctuations in the number of trimers.
We did so by studying a TN perturbation that represents a diluted tRVB state on the square lattice and showing that it hosts a wide topologically ordered phase in the state phase diagram.
Finally, we considered a simple model Hamiltonian on the square lattice that exhibits signatures of a tRVB-like phase, where the ground state is well approximated by the $\mathbb{Z}_3$ topologically ordered diluted TN perturbation previously studied.
We provided an implementation of a very similar model by exploiting the blockade effect in Rydberg atom arrays, and show that hallmarks of a $\mathbb{Z}_3$ quantum spin liquid can be observed in non-adiabatic dynamical preparation schemes.

Our findings open future directions for the quantum simulation of topological phases of matter.
The necessary condition for having $\mathbb{Z}_3$ topological order that we formulated depends solely on the geometry of the model and can therefore guide the search for quantum spin liquids in various experimental implementations, including---but not limited to---Rydberg atom arrays.
In this respect, it would be interesting to study more extensively the realization of tRVB states both as ground states of realistic Hamiltonians and as dynamically-prepared non equilibrium states.
In addition, our approach can be naturally extended from trimer- to polymer-RVB states, which can support the emergence of $\mathbb{Z}_n$ topological order.
A systematic study of such states can similarly be performed efficiently with tensor network methods and is left for future work.
Finally, an interesting direction is the related problem of quantum spin liquid phases in $\mathrm{SU}(3)$ and $\mathrm{SU}(N)$ symmetric models.
In certain models, trimers (polymers) can be interpreted as simplified versions of $\mathrm{SU}(3)$ [$\mathrm{SU}(N)$] spin singlets; it remains an open question to what extent this interpretation can be used to infer the properties of RVB states of singlets.

\begin{acknowledgments}
    We acknowledge inspiring discussions with Norbert Schuch, Juraj Hasik, Andrej Gendiar and Olexei Motrunich.
    TN calculations were performed by resolving symmetries at the tensor level with the TensorKit.jl package~\cite{tensorkit}.
    This work has been supported by the European Research Council (ERC) under the European Union’s Horizon 2020 research and innovation programme through the ERC-CoG \mbox{SEQUAM} (Grant No.~863476), the ERC-CoG QSIMCORR (Grant No.~771891), and ERC-StG \mbox{QARA} (Grant No.~101041435), as well as the Deutsche Forschungsgemeinschaft (DFG, German Research Foundation) under Germany's Excellence Strategy  (Grant No. ~EXC-2111 -- 390814868).
\end{acknowledgments}
\bibliography{bibliography}

\appendix
\section{Details on the tensor-network representations}
\label{app:details}

\subsection{tRVB model on the square lattice}
\label{app:details:square}
\subsubsection{Symmetries}
The tensor defined in \cref{eq:tensor_square} enjoys a reflection symmetry along the diagonal, which can be exploited in the CTMRG algorithm.
The corresponding transfer operator is not self-adjoint, which requires one to compute both fixed points for the VUMPS algorithm.
However, the fixed point in one direction can be readily converted into the the fixed point into the other direction, since
\begin{equation}
    \includescaledgraphics{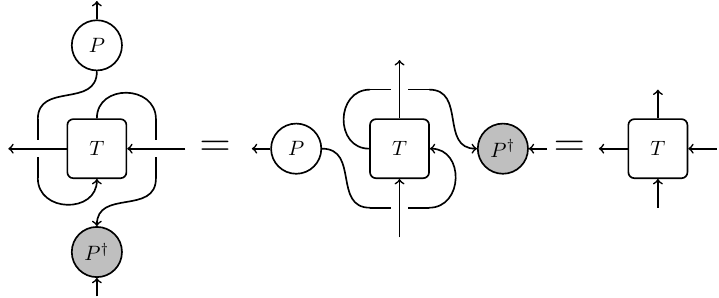}\, ,
\end{equation}
where $P$ is the operator that permutes the $\pm 1$ indices.

At $\theta=\pi/2$, a representation of the U$(1)\times$U$(1)$ symmetry at the level of a single tensor can be obtained as follows.
The symmetry transformation on a site of type ``A'' (and similarly for ``B'' and ``C'', with a cyclic permutation of the indices) acts as
\begin{equation}
    \includescaledgraphics{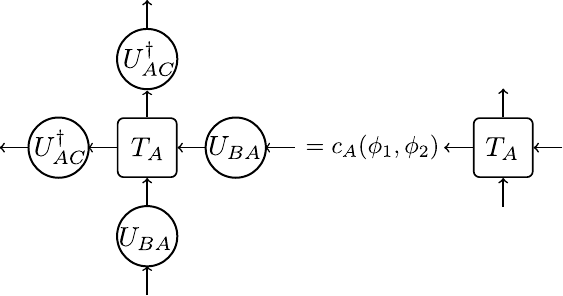}\, .
\end{equation}
where $U_{BA}=\exp\big(i\phi_1G_{BA}^{(1)}+i\phi_2G_{BA}^{(2)}\big)$ (and analogous definitions for the links $AC$, $CB$).
The generators of the two symmetries in the basis $\{0,+1,-1\}$ of the virtual indices are
\begin{subequations}
    \begin{align}
    G_{BA}^{(1)} &= -G_{AC}^{(2)} = \begin{pmatrix}
    0 & 0& 0\\
    0 & 1 & 0\\
    0 & 0 & 1\\
    \end{pmatrix},
    \\
    G_{CB}^{(1)} &= -G_{BA}^{(2)} = \begin{pmatrix}
    0 & 0& 0\\
    0 & -1 & 0\\
    0 & 0 & 0\\
    \end{pmatrix},
    \\
    G_{AC}^{(1)} &= -G_{CB}^{(2)} = \begin{pmatrix}
    0 & 0& 0\\
    0 & 0 & 0\\
    0 & 0 & -1\\
    \end{pmatrix}.
    \end{align}
\end{subequations}
The phases acquired by the tensor read
\begin{subequations}
    \begin{align}
    c_A(\phi_1, \phi_2) &= e^{i\phi_1+i\phi_2},\\
    c_B(\phi_1, \phi_2) &= e^{-i\phi_1},\\
    c_C(\phi_1, \phi_2) &= e^{-i\phi_2}.
    \end{align}
\end{subequations}
To obtain a translationally-invariant and neutral tensor representation, one must block a $3 \times 3$ patch of elementary tensors.

\subsubsection{Diluted tRVB PEPS}
The PEPS representation for the diluted tRVB in \cref{eq:diluted_trvb} is constructed by introducing two species of trimers; one that appear on the physical layer and ones that do not.
Graphically, we have the following construction:
\begin{equation}
    \ket{\Phi} = \hspace{-2em}\includescaledgraphics{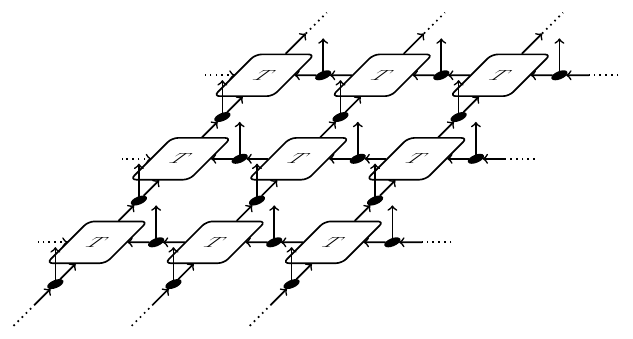}.
\end{equation}
We label the indices of the tensor $\tilde{T}$ with $\{0, +1, -1, +\tilde{1}, -\tilde{1}\}$.
We now have two pairs of uncoupled charges, each of which have the similar diagrams as \cref{eq:tensor_square}.
In order to preserve the correspondence with the partition function in \cref{eq:partition}, we must also take the square root of each weight.
The black dot on each link represents a projector, which signals the presence of one type of trimer on the physical layer, and gives a fugacity contribution $z$ to the second type
\begin{equation}
    \includescaledgraphics{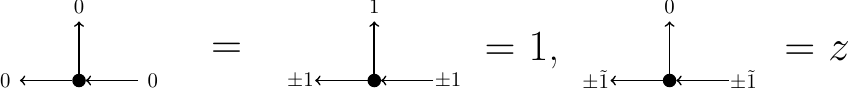}\, .
\end{equation}
The corresponding PEPS tensor, formed by contracting one $\tilde{T}$ and two projectors, has bond dimension $D = 5$.
It should be noted that the double-layer tensor---coming from the contraction of the norm $\braket{\Phi|\Phi}$---can be reduced to dimension 9 (instead of the naïvely expected $5^2$).
We observe this by performing Gaussian elimination on the tensor, which is exact up to numerical precision.

\subsubsection{Restricted tRVB model}
\begin{figure}[ht]
    \includegraphics[width=0.65\columnwidth]{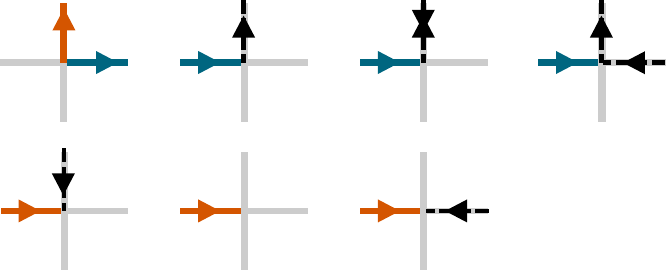}
    \caption{Allowed diagrams at each vertex for the restricted tRVB model, up to rotations of $\pi/2$.}
    \label{fig:trvb_restricted}
\end{figure}

The mapping to Rydberg atoms in \cref{sec:ryd} suggests the ansatz state in which we restrict the trimer configurations to not have any ``wedged'' configurations, as shown in \cref{fig:implementation}.
We start from the arrow representation in \cref{fig:z3gauss:validconfigs}, but we introduce new color labels
\begin{equation}
    \includegraphics[width=0.65\columnwidth]{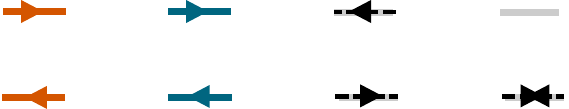} \, .
\end{equation}
Referring to \cref{fig:trvb_restricted}, we start by labeling the two outgoing arrows of a bent trimer with different colors (blue and orange).
To enforce the additional constraint, we introduce a dashed arrow which continues anticlockwise from the blue arrow.
Notice that the arrow does not convey any charge information, as all dashed lines correspond to a $\mathbb{Z}_3$-charge 0.
A vertex with the middle of the trimer cannot couple to it directly since it requires two gray lines.
The other diagrams then account for the possible ways of closing the dashed lines.
Accounting for rotations, in total there are 28 distinct diagrams.

The construction of the tensor network from \cref{fig:trvb_restricted} is straightforward, and is similar to \cref{eq:tensor_square}.
We must however remember that the mapping from the arrow representation to the index of the tensor is different if a leg of the tensor is ingoing or outgoing.
Each leg of the double-layer tensor is then eight-dimensional.

\subsection{tRVB model on the honeycomb lattice}
\label{app:details:honeycomb}
On the honeycomb lattice there is only one type of trimer.
On each vertex of the lattice we can place the tensor
\begin{equation}
    \left.
    \includescaledgraphics{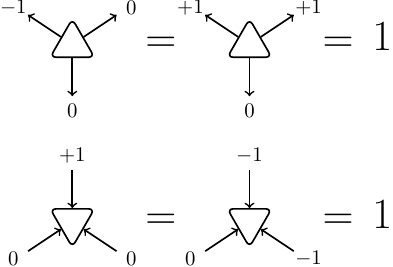} \ \right\} \mbox{$\forall$ rotations}
    \label{eq:tensor_honeycomb}
\end{equation}
to obtain the corresponding partition function.
To convert the problem into a TN on the square lattice, we define the tensor
\begin{equation}
    \includescaledgraphics{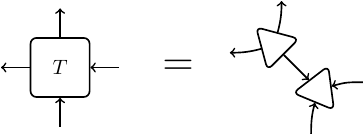} \, .
\end{equation}
Similarly to the case of square lattice in \cref{eq:symmetry}, this tensor obeys a similar transformation, $(\sigma^\dag \otimes \sigma^\dag \otimes \sigma \otimes \sigma) T = \omega^2 T$.

\subsection{tRVB model on the triangular lattice}
\label{app:details:triangular}
On the triangular lattice an efficient TN representation can be found in the dual-lattice picture, where a site is defined on each triangular face, similarly to the way in which one would propose a Rydberg implementation.
In this picture, only one site around each original vertex of the triangular lattice can be occupied.
Correspondingly, in the TN picture we define a $\delta$-tensor on each face, representing the site
\begin{equation}
    \includescaledgraphics{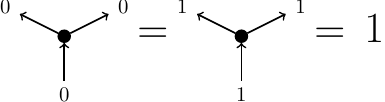} \, ,
\end{equation}
which is connected to three rank-6 constraint tensors
\begin{equation}
    \left.
    \includescaledgraphics{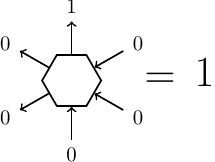} \ \right\} \mbox{$\forall$ rotations} \, .
\end{equation}
We can then bring the problem back to a TN contraction on the square lattice by first defining a decomposition of the constraint tensor
\begin{equation}
    \includescaledgraphics{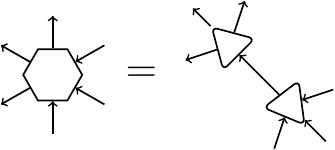} \, ,
\end{equation}
and performing the contraction
\begin{equation}
    \includescaledgraphics{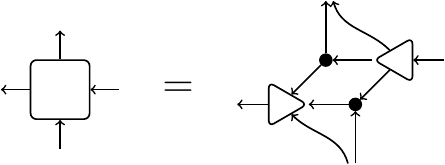} \, .
\end{equation}

\section{Proof of \texorpdfstring{$\mathrm{U}(1)\times\mathrm{U}(1)$}{U₁×U₁} gauge symmetry}
\label{app:details:indep}
In \cref{sec:u1u1} we showed that two $\mathrm{U}(1)$ flux conservation laws could be defined for a model with straight trimers on the square lattice. To prove that a $\mathrm{U}(1)\times\mathrm{U}(1)$ gauge theory emerges in the model we have to further prove that the two conservation laws are independent. We here prove it by showing that the number of sectors for a semi-infinite cylinder of circumference $L$ is at least $O(L^2)$.

\begin{figure}[b]
    \includegraphics[width=\columnwidth]{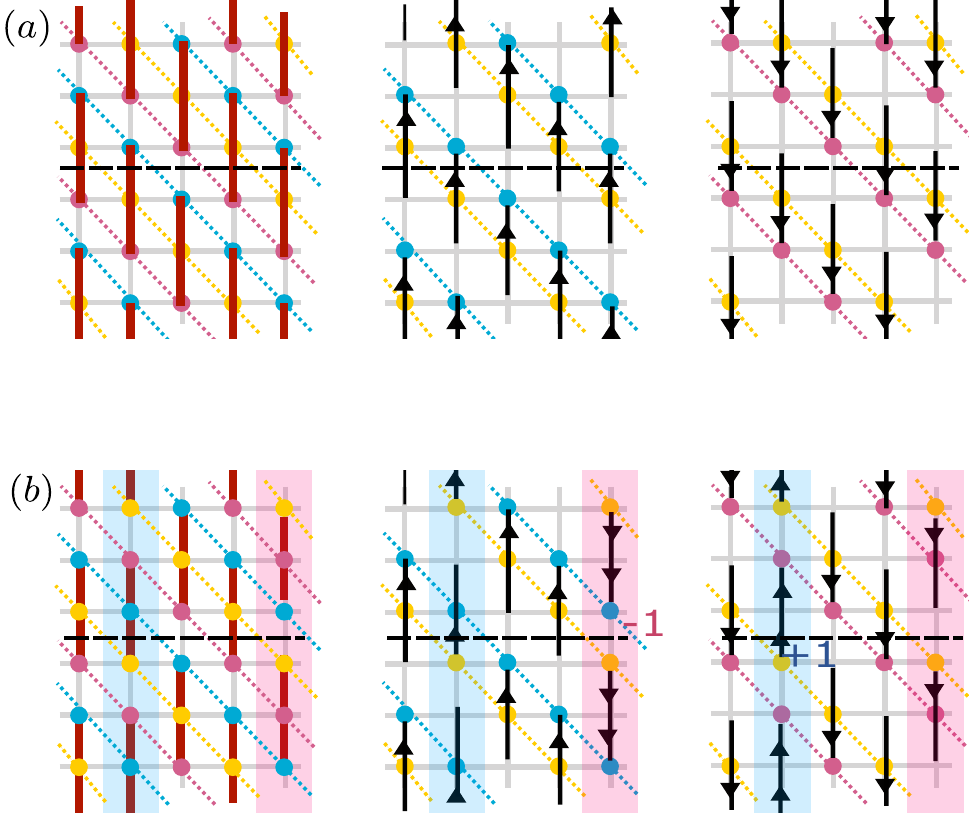}
    \vspace*{-3mm}
    \phantomsubfloat{\label{fig:proof1}}
    \phantomsubfloat{\label{fig:proof2}}
    \caption{%
        (a)~A columnar configuration of straight trimers and the corresponding electric field configurations. The dashed-black line is the cut that defines the semi-infinite cylinder. $\Phi_1$ and $\Phi_2$ are the fluxes of the two electric field across the cut.
        (b)~Each column can shifted up (in blue) or down (in red) of one lattice site to obtain a new configuration of trimers. Shifting a single column changes only one of the two fluxes.
    }
    \label{fig:proof}
\end{figure}

Let us consider the trimer configuration in \cref{fig:proof1}. To obtain a cylinder, we impose periodic boundary condition in the horizontal direction and call $L$ the circumference (where $L$ is an integer multiple of 3), while the vertical direction extends indefinitely. We perform a horizontal cut [dashed black line in \cref{fig:proof1}] and we compute the two electric fluxes along the cut. We assign a positive sign to the electric fields pointing upwards along the cut. The configuration in \cref{fig:proof1} has fluxes $\Phi_1=+L/3$ and $\Phi_2=-L/3$. From this configuration, we can systematically construct configurations belonging to other sectors by shifting some columns in the vertical direction [\cref{fig:proof2}]: a column can be shifted up (highlighted in blue) or down (in red) of one site; these are the only two relevant cases, because of the $\mathbb{Z}_3$ symmetry (i.e., the columns are invariant under a 3-site translation). Comparing \cref{fig:proof1} and \cref{fig:proof2} we see that (i) shifting one column up gives $\Phi_1\rightarrow \Phi_1$, $\Phi_2\rightarrow \Phi_2+1$, (ii) shifting one column down results in $\Phi_1\rightarrow \Phi_1-1$, $\Phi_2\rightarrow \Phi_2$. It can be easily verified that this result is valid for every column. Therefore, by shifting $n_\uparrow$ columns up and $n_\downarrow$ columns down, we obtain a configuration that belongs to the sector with electric fluxes $\Phi_1=L/3-n_\downarrow, \Phi_2=-L/3+n_\uparrow$. Because we can choose any pair of integers $n_\downarrow, n_\uparrow\ge 0$ with $n_\downarrow+n_\uparrow\le L$, the number of different sectors that we obtain is $(L+1)(L+2)/2$. This proves that the number of sectors scales proportionally to $L^2$, meaning that the two emergent U$(1)$ gauge symmetries are independent. A similar argument can be used to prove the emergence of a $\mathrm{U}(1)\times\mathrm{U}(1)$ symmetry for other tripartite trimer models.

\end{document}